\documentclass[sigconf,screen]{acmart}

\usepackage[T1]{fontenc}

\usepackage{microtype}
\usepackage{xspace}
\usepackage{color}
\usepackage{xcolor}
\usepackage{enumitem}
\usepackage{balance}
\usepackage{multirow}
\usepackage{tabularx}
\usepackage{float}
\usepackage{algorithm}
\usepackage{booktabs}
\usepackage{threeparttable}
\usepackage[noend]{algpseudocode}
\DeclareCaptionSubType*{algorithm}
\usepackage{tikz}
\usetikzlibrary{positioning, arrows.meta, fit, backgrounds, calc}
\usepackage{graphicx}
\usepackage{subcaption}
\usepackage{array}

\usepackage{fancyvrb}
\input{listings/gen/pygments-style.tex}

\usepackage{hyperref}
\usepackage[capitalise,nameinlink]{cleveref}

\crefformat{section}{#2\S{}#1#3}

\usepackage{ragged2e}

\newcommand{\PP}[1]{
\noindent{\bf \IfEndWith{#1}{.}{#1}{#1.}}
}

\newcommand{\spell}[1]{%
  \ifcase\numexpr#1\relax
    zero\or one\or two\or three\or four\or five\or
    six\or seven\or eight\or nine\or ten\else\number\numexpr#1\relax\fi}
\newcommand{\Spell}[1]{%
  \ifcase\numexpr#1\relax
    Zero\or One\or Two\or Three\or Four\or Five\or
    Six\or Seven\or Eight\or Nine\or Ten\else\number\numexpr#1\relax\fi}

\newcommand{\framework}{\textsc{SpecTrum}\xspace}
\newcommand{\spectec}{SpecTec\xspace}
\newcommand{\mech}{\textsc{Consensus-SpecTec}\xspace}
\newcommand{\cspec}{\textsf{consensus-spec}\xspace}
\newcommand{\ctests}{\textsf{spectests}\xspace}
\newcommand{\ctrlrand}{\textsf{Random}\xspace}
\newcommand{\ctrlext}{\textsf{Extreme}\xspace}

\definecolor{codebinding}{HTML}{292988}
\definecolor{codekeyword}{HTML}{6A6AD9}
\definecolor{codeliteral}{HTML}{A95FC7}
\definecolor{codefield}{HTML}{404040}
\definecolor{codeop}{HTML}{737373}

\newcommand\specplain[1]{{\normalfont\texttt{\small{#1}}}}
\newcommand\specfn[1]{{\normalfont\texttt{\small{\color{codebinding}#1}}}}
\newcommand\specfield[1]{{\normalfont\texttt{\small{\color{codefield}#1}}}}
\newcommand\speckw[1]{{\normalfont\texttt{\small{\bfseries\color{codekeyword}#1}}}}
\newcommand\specop[1]{{\normalfont\texttt{\small{\color{codeop}#1}}}}
\newcommand\specnum[1]{{\normalfont\texttt{\small{\color{codeliteral}#1}}}}
\newcommand\code[1]{{\normalfont\texttt{\small{#1}}}}

\newcommand\cfn[1]{{\color{codebinding}#1}}
\newcommand\cfield[1]{{\color{codefield}#1}}
\newcommand\cop[1]{{\color{codeop}#1}}

\usepackage{amssymb}

\newsavebox{\implbox}

\newcommand{\implres}[3]{%
  \sbox{\implbox}{#1}%
  \makebox[\dimexpr\wd\implbox+0.75em\relax][l]{%
    \ooalign{%
      \usebox{\implbox}\cr
      \smash{\raisebox{-0.95ex}{\makebox[\wd\implbox][c]{#3}}}\cr
    }%
    \hspace{0.15em}%
    \makebox[0.60em][c]{#2}%
  }%
}

\newcommand{\resfail}[1]{\implres{#1}{$\times$}{}}
\newcommand{\ressuc}[1]{\implres{#1}{$\checkmark$}{}}
\newcommand{\rescrash}[1]{\implres{#1}{$\blacktriangle$}{}}
\newcommand{\suc}{\ensuremath{\checkmark}}
\newcommand{\fal}{\ensuremath{\times}}
\newcommand{\cra}{\ensuremath{\blacktriangle}}

\newcommand{\resone}[1]{%
  \implres{#1}{$\checkmark$}{\scalebox{0.55}{$\bullet$}}%
}

\newcommand{\restwo}[1]{%
  \implres{#1}{$\checkmark$}{\scalebox{0.55}{$\bullet\;\bullet$}}%
}

\newif\ifshowchanges
\showchangesfalse  

\definecolor{revisionblue}{HTML}{0057B8}

\newcommand{\numValidationTests}{910}          
\newcommand{\numStateTransitionTests}{581}     
\newcommand{\numUnitTests}{329}                

\newcommand{\numTotalIfPremises}{392}          
\newcommand{\numLOC}{2{,}632}
\newcommand{\numRel}{65}
\newcommand{\numFunc}{138}
\newcommand{\numInsertedChecks}{104}           
\newcommand{\numBoundsChecks}{23} 

\newcommand{\numOverOrUnderflowChecks}{80}
\newcommand{\numDivisorChecks}{1}              

\newcommand{\numTautologyPremises}{56}         
\newcommand{\numClosingBranchPremises}{55}     
\newcommand{\numUnfalsifiablePremises}{111}    

\newcommand{\numTotalFalsifiablePremises}{281}
\newcommand{\numInsertedFalsifiable}{81}      
\newcommand{\numBaselineFalsified}{130}         
\newcommand{\numNewlyFalsifiablePremises}{151}  
\newcommand{\numUnattemptedPremises}{3}         
\newcommand{\numTargetPremises}{137}

\newcommand{\numGeneratedTests}{21{,}444}
\newcommand{\numSeeds}{59}                     

\newcommand{\numDivergences}{27}               
\newcommand{\numDivergenceCategories}{6}       
\newcommand{\numImplicitDivergences}{22}       
\newcommand{\caseOverflow}{17--21}             

\newcommand{\numNewFalsified}{126}             
\newcommand{\numNewFalsifiedInserted}{65}      
\newcommand{\numNewFalsifiedExplicit}{58}      
\newcommand{\numNewFalsifiedSynth}{3}          
\newcommand{\numCombinedFalsified}{256}        
\newcommand{\pctBaselinePremiseCoverage}{46.3} 
\newcommand{\pctCombinedPremiseCoverage}{91.1} 
\newcommand{\numRemaining}{25}                 
\newcommand{\numLimitation}{22}                
\newcommand{\numCoordinatedLimitation}{19}
\newcommand{\numProvenanceLimitation}{3}

\newcommand{\numCtrlTotal}{\numTotalFalsifiablePremises}
\newcommand{\numCtrlBaseline}{\numBaselineFalsified}
\newcommand{\numCtrlExtreme}{187}
\newcommand{\numCtrlRandom}{193}
\newcommand{\numCtrlSpecTrum}{\numCombinedFalsified}
\newcommand{\numCtrlExtremeDelta}{57}
\newcommand{\numCtrlRandomDelta}{63}
\newcommand{\numCtrlSpecTrumDelta}{\numNewFalsified}
\newcommand{\numExtremeKinds}{16}              
\newcommand{\numRandomKinds}{12}               
\newcommand{\numExtremeDivergences}{11}        
\newcommand{\numRandomDivergences}{7}
\newcommand{\numRandomControlOnly}{5}          
\newcommand{\numControlUnion}{12}              
\newcommand{\numExtremeClassA}{4}              
\newcommand{\numRandomClassA}{0}               
\newcommand{\numClassA}{4}                     
\newcommand{\numClassB}{23}

\newcommand{\numDenebLineChanges}{26}          
\newcommand{\numDenebFilesChanged}{10}
\newcommand{\numDenebFilesTotal}{22}

\newcommand{\numDenebTestCases}{21{,}825}      
\newcommand{\numDenebTotalIfPremises}{388}
\newcommand{\numDenebBaselineFalsified}{131}
\newcommand{\numDenebCombinedFalsified}{261}
\newcommand{\numDenebNewFalsified}{130}        
\newcommand{\numDenebTotalFalsifiablePremises}{280}   
\newcommand{\pctDenebCombinedPremiseCoverage}{93.2}   

\copyrightyear{2026}
\acmYear{2026}
\setcopyright{cc}
\setcctype{by}
\acmConference[ASE '26]{Proceedings of the 41st IEEE/ACM International Conference on Automated Software Engineering}{October 12--16, 2026}{Munich, Germany}
\acmBooktitle{Proceedings of the 41st IEEE/ACM International Conference on Automated Software Engineering (ASE '26), October 12--16, 2026, Munich, Germany}
\acmDOI{10.1145/3832783.3834406}
\acmISBN{979-8-4007-2882-2/2026/10}



\begin{document}

\title{\framework{}: Specification-Guided Differential Fuzzing for Ethereum Consensus Clients}

\author{Seokhun Jeong}
\orcid{0009-0008-4273-3508}
\affiliation{%
  \institution{KAIST}
  \city{Daejeon}
  \country{Republic of Korea}
}
\email{sraccoon@kaist.ac.kr}

\author{Gyeongmin Dan}
\orcid{0009-0001-9502-6320}
\affiliation{%
  \institution{Sungkyunkwan University}
  \city{Suwon}
  \country{Republic of Korea}
}
\email{albert1121@skku.edu}

\author{Sukyoung Ryu}
\orcid{0000-0002-0019-9772}
\affiliation{%
  \institution{KAIST}
  \city{Daejeon}
  \country{Republic of Korea}
}
\email{sryu.cs@kaist.ac.kr}

\author{Sungjae Hwang}
\authornote{Corresponding author.}
\orcid{0000-0001-5386-2411}
\affiliation{%
  \institution{Sungkyunkwan University}
  \city{Suwon}
  \country{Republic of Korea}
}
\email{sungjaeh@skku.edu}

\begin{abstract}
Ethereum's consensus safety relies on independent consensus client implementations agreeing on every state transition.
When they diverge due to implementation errors, the network can fork, finality can stall, and severe attacks are possible.
To prevent such \textit{consensus divergences},
Ethereum provides a Python reference implementation (\cspec{}), which acts as a specification,
and a hand-crafted official test suite (\ctests{}).
However, as an executable implementation,
Ethereum's specification defines validity implicitly through runtime behavior.
As a result, it lacks a systematic way to ensure that all validity conditions are thoroughly evaluated.

We present \framework{}, a framework that addresses this problem in three stages.
First, we introduce \mech{},
a mechanized specification of the Ethereum consensus algorithm,
which makes validity conditions explicit as \textit{if-premises}.
Second, we define \textit{premise coverage},
a metric that measures which if-premises are evaluated to true and false across \ctests{}.
Third, we develop a specification-based test generator
that extracts constraints on premises not evaluated to false
by \ctests{} and generates inputs to evaluate them.
Applying \framework{} to five major Ethereum consensus clients,
we identify \numDivergences{} cross-client divergence cases,
\numImplicitDivergences{} of which cannot be found
without the premises inserted in our mechanization.
All \numDivergences{} cases reproduce across fork versions,
and extending the mechanized specification to a new fork
takes modest effort proportional to the specification difference.
\end{abstract}

\maketitle

\section{Introduction}\label{sec:intro}
Ethereum's safety depends on independent consensus clients computing the same 
state transition for the same pre-state and block.
Ethereum promotes client diversity to reduce the risk of single-implementation failures~\cite{clientdiversity}:
major clients include
Lighthouse (Rust)~\cite{Lighthouse}, Lodestar (TypeScript)~\cite{Lodestar},
Nimbus (Nim)~\cite{Nimbus}, Prysm (Go)~\cite{Prysm}, and Teku (Java)~\cite{Teku}.
Diversity improves resilience, but differences in how clients interpret the protocol
can cause inconsistent state-transition outcomes.
Such inconsistencies may lead to incorrect validator penalties,
chain splits, or stalled finality, threatening network reliability~\cite{prysmfinality2023,beaconfuzz,yang2021}.

The primary conformance baseline is the official Ethereum consensus specification, 
\cspec{}~\cite{ethereumConsensusSpec}, together with the accompanying hand-written specification tests (\ctests{})~\cite{consensusSpecTests},
which clients use to check conformance~\cite{prysm12884,prysmtracking,lighthousetracking}.
As executable Python, \cspec{} is \emph{complete}:
every transition either produces a post-state or raises an exception,
as the specification states~\cite{consensusSpecStateTransition}:
\begin{quote}
State transitions that trigger an unhandled exception
(e.g.\ a failed \specplain{assert} or an out-of-range list access) are considered invalid.
State transitions that cause a \specplain{uint64} overflow or underflow are also 
considered invalid.
\end{quote}
This convention, however, defines invalidity only through runtime behavior,
rather than as explicit conditions.
In addition,
the checks are scattered across overlapping conditional statements
and in-place state mutations.
A developer porting \cspec{} to other languages
must recover these conditions from the code,
not from a list of validity rules.
We call such conditions \emph{implicit}:
not absent or ambiguous, but hidden behind implementation details
such as mutation, condition fallthrough, or runtime exceptions.
Such symptoms are not unique to Ethereum: specifications are often given as
reference implementations that encode validity implicitly.
As we show in \cref{sec:rq2}, this implicitness hides these conditions from code-level coverage,
and leads to under-testing.
This is a significant limitation,
because consensus bugs arise from such validity conditions,
particularly boundary cases~\cite{ethspec1701,beaconfuzz,yang2021}.
Existing differential fuzzers~\cite{beaconfuzz,yang2021,kim2025fork,ma2023loki} cannot sufficiently cover such boundaries,
because they select mutation targets without reference to the specification's
validity conditions.

We address this problem with \framework{}, a specification-guided differential testing 
framework for Ethereum consensus clients. The key component of \framework{} is \mech{},
a mechanized specification of the Ethereum consensus algorithm.
It explicitly specifies the mutation semantics, 
branching structure, and implicit exception boundaries as \emph{if-premises}. 
On top of this mechanized specification, we define \textit{premise coverage},
a structural coverage metric that measures
whether each failure-relevant premise is evaluated to both true and false.
Because these conditions are now explicit, premise coverage finds
validity conditions that \ctests{} under-executes
and code-level coverage cannot see.
\framework{} uses \ctests{} as seeds, tracks how 
input values (e.g., states) are derived during execution, extracts 
constraints from under-executed premises, and generates tests 
that execute those premises or target their boundary values.
The generated tests are executed across independent clients, 
and disagreements in outcome or post-state are reported as divergences.

Our evaluation shows that \framework{} is both effective and practical. 
We first validate \mech{} against \numValidationTests{} official tests, including \numStateTransitionTests{} state-transition tests
and \numUnitTests{} unit tests, and obtain full agreement with \cspec{}.
We then apply \framework{} to the five major Ethereum consensus clients
and identify \numDivergences{} cross-client divergences across
\spell{\numDivergenceCategories} root-cause categories, leading to consensus
failures, liveness failures, and client crashes.
Some divergences are silent: clients accept the same input
yet compute different post-states, undetectable by accept/reject testing.
We have reported all the identified divergences to the Ethereum Protocol Bug Bounty 
Program (\code{bounty@ethereum.org}) and they are awaiting triage.

The technical contributions of this paper include the following:
\begin{itemize}
\item \textbf{\mech{}}, a mechanized specification of the Ethereum consensus algorithm,
which makes \cspec{}'s implicit validity conditions explicit as if-premises.
It is the first application of the \spectec approach~\cite{spectec,p4spectec} to
non-programming-language software.
\item \textbf{Premise coverage}, a structural coverage metric defined over if-premises,
which identifies \numNewlyFalsifiablePremises{} conditions not covered by \ctests{},
\numInsertedFalsifiable{} of which are inserted.
\item \textbf{A specification-guided test generator}
that traces input provenance through \mech{},
extracts constraints to cover the target conditions,
and samples values, raising premise coverage
from \pctBaselinePremiseCoverage{}\% to \pctCombinedPremiseCoverage{}\%.
\item \textbf{A comprehensive empirical evaluation}
demonstrating \framework{}'s effectiveness and practicality through
improved premise coverage, \numDivergences{} cross-client divergence
cases across \spell{\numDivergenceCategories} root-cause categories, superiority over
unguided mutation, and transfer across protocol forks.
\end{itemize}

\section{Background}\label{sec:background}
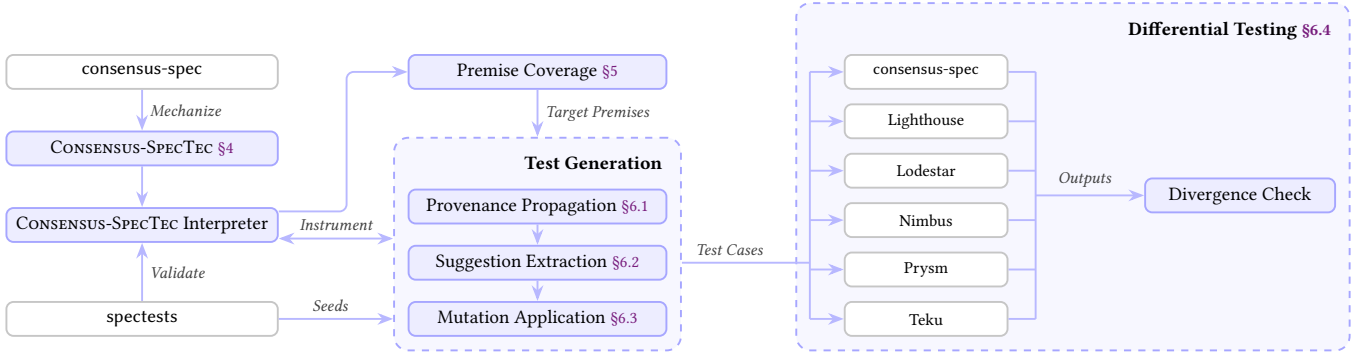
\begin{figure*}[t]
\centering
\resizebox{\textwidth}{!}{%
\begin{tikzpicture}[
    >=Stealth,
    box/.style={
        rectangle, rounded corners=3pt,
        minimum height=0.5cm, minimum width=3.8cm,
        font=\small, align=center, thick
    },
    procbox/.style   = {box, fill=blue!7, draw=blue!35},
    inputbox/.style  = {box, fill=white, draw=gray!45},
    col1/.style      = {minimum width=4.0cm},
    clientbox/.style = {box, fill=white, draw=gray!45, minimum width=2.4cm, minimum height=0.5cm, font=\footnotesize},
    outerbox/.style  = {draw=blue!35, dashed, thick, rounded corners=6pt, fill=blue!3, inner xsep=0.2cm, inner ysep=0.2cm},
    arr/.style  = {->, thick, blue!28},
    darr/.style = {<->, thick, blue!28},
    lbl/.style  = {font=\footnotesize\itshape, text=black!75},
]

\node[inputbox, col1] (conspec) {\cspec{}};
\node[procbox, col1, below=0.6cm of conspec] (mech) {\mech{} {\footnotesize\textcolor{black!45}{\cref{sec:spec}}}};
\node[procbox, col1, below=0.6cm of mech] (interp) {\mech{} Interpreter};

\node[procbox, right=1.9cm of conspec.north east, anchor=north west] (premcov) {Premise Coverage {\footnotesize\textcolor{black!45}{\cref{sec:coverage}}}};

\node[clientbox, right=2.6cm of premcov.north east, anchor=north west] (cs) {\cspec{}};
\node[clientbox, below=0.2cm of cs] (lh) {Lighthouse};
\node[clientbox, below=0.2cm of lh] (lo) {Lodestar};
\node[clientbox, below=0.2cm of lo] (ni) {Nimbus};
\node[clientbox, below=0.2cm of ni] (pr) {Prysm};
\node[clientbox, below=0.2cm of pr] (te) {Teku};

\node[inputbox, col1, anchor=south] (spectests) at (conspec.south |- te.south) {\ctests{}};

\coordinate (tg_x) at ($(interp.north east) + (1.9cm, 0)$);
\node[procbox, anchor=south west] (mut) at (tg_x |- te.south) {Mutation Application {\footnotesize\textcolor{black!45}{\cref{sec:mutation}}}};
\node[procbox, above=0.3cm of mut] (sugg) {Suggestion Extraction {\footnotesize\textcolor{black!45}{\cref{sec:suggestion}}}};
\node[procbox, above=0.3cm of sugg] (prov) {Provenance Propagation {\footnotesize\textcolor{black!45}{\cref{sec:provenance}}}};

\coordinate (center_clients) at ($(cs.north east)!0.5!(te.south east)$);
\node[procbox, right=2.0cm of center_clients, anchor=west, minimum width=2.8cm] (divcheck) {Divergence Check};

\coordinate (left_spine) at ($(cs.west) + (-0.5cm, 0)$);
\draw[thick, blue!28, rounded corners=3pt] (left_spine |- cs.west) -- (left_spine |- te.west);
\foreach \cl in {cs, lh, lo, ni, pr, te} {
    \draw[arr, thick, blue!28] (left_spine |- \cl.west) -- (\cl.west);
}

\coordinate (right_spine) at ($(cs.east) + (0.4cm, 0)$);
\foreach \cl in {cs, lh, lo, ni, pr, te} {
    \draw[thick, blue!28] (\cl.east) -- (right_spine |- \cl.east);
}
\draw[thick, blue!28, rounded corners=3pt] (right_spine |- cs.east) -- (right_spine |- te.east);

\begin{scope}[on background layer]
    \coordinate (phase3_top) at ($(prov.north) + (0, 0.55cm)$);
    \node[outerbox, fit=(prov)(sugg)(mut)(phase3_top)] (phase3box) {};

    \coordinate (right_spine_top) at ($(right_spine |- cs.north) + (0, 0.55cm)$);
    \coordinate (left_spine_bottom) at (left_spine |- te.south);
    \node[outerbox, fit=(cs)(te)(divcheck)(left_spine_bottom)(right_spine_top)] (testbox) {};
\end{scope}

\node[font=\small\bfseries, anchor=north east, inner sep=8pt]
    at (phase3box.north east) (testgen_title)
    {Test Generation};
\node[font=\small\bfseries, anchor=north east, inner sep=8pt]
    at (testbox.north east) (difftest_title)
    {Differential Testing {\footnotesize\textcolor{black!45}{\cref{sec:diff_testing}}}};

\draw[arr] (conspec) -- node[right, lbl] {Mechanize} (mech);
\draw[arr] (mech) -- (interp);
\draw[arr] (spectests) -- node[right,lbl] {Validate} (interp); 

\draw[arr] (prov) -- (sugg);
\draw[arr] (sugg) -- (mut);

\draw[arr] (premcov.south) -- node[right, lbl] {Target Premises} (premcov.south |- phase3box.north);

\coordinate (interp_up) at ($(interp.east) + (0, 0.2cm)$);
\draw[arr, rounded corners=3pt] (interp_up) -- ++(1.0, 0) |- (premcov.west);

\coordinate (interp_down) at ($(interp.east) + (0, -0.2cm)$);
\draw[darr] (interp_down) -- node[above, lbl] {Instrument} (phase3box.west |- interp_down);

\draw[arr] (spectests.east) -- node[above, pos=0.45, lbl] {Seeds} (phase3box.west |- spectests.east);

\coordinate (phase3_out) at ($(phase3box.east |- sugg.east)$);
\draw[thick, blue!28] (phase3_out) -- node[above, pos=0.45, xshift=-0.15cm, lbl] {Test Cases} (left_spine |- phase3_out);

\coordinate (divcheck_in) at ($(right_spine)!0.5!(right_spine |- cs.east)$);
\draw[arr, thick, blue!28] (divcheck_in |- divcheck.west) -- node[above, pos=0.45, lbl] {Outputs} (divcheck.west);

\end{tikzpicture}%
}
\caption{Overview of the \framework{} framework.
}
\label{fig:overview}
\end{figure*}


\subsection{Ethereum State Transition}
In Ethereum's Proof-of-Stake protocol (Gasper~\cite{buterin2020combining}),
\textit{validators} are network participants who stake ETH to propose and attest to blocks.
The protocol divides time into fixed-length \textit{slots},
grouped into \textit{epochs} of 32 slots each.
In each slot, one validator is chosen as the \textit{proposer} and submits a \specplain{BeaconBlock},
which packages new transactions, attestations, and validator operations.
All other validators are \textit{attesters} who validate the proposed block.
Each validator runs a \textit{consensus client}:
an implementation that handles consensus logic,
including block proposal, validation, and fork choice.
These operations update the \specplain{BeaconState}, a persistent data structure
that stores validator registries, balances, committee assignments, and finalized checkpoints.
States and blocks are serialized with SSZ, the protocol's encoding.
Each consensus client $C$ implements the \emph{state-transition function} $f_C$,
defined by the Ethereum consensus specification.
This function takes the current \specplain{BeaconState} $S$
and an incoming \specplain{BeaconBlock} $B$
and returns $f_C(S, B) = S'$, the updated state,
or $f_C(S, B) = \bot$ if the transition is rejected.

This function is the target of our testing framework.
It proceeds in two phases, as shown in the call tree of
\specfn{state\_transition}\specplain{\cop{(}\cfield{state}\cop{,} \cfield{block}\cop{)}} below:

\par\medskip
\noindent\hspace*{1em}%
\begin{tikzpicture}[
  font=\ttfamily\footnotesize,
  treenode/.style={anchor=base west, inner sep=0pt, outer sep=0pt},
  guide/.style={gray, line width=0.4pt},
  x=1.4em, y=0.93em,
]
  \node[treenode]       (a) at (0,0)   {\cfn{state\_transition}\cop{(}\cfield{state}\cop{,} \cfield{block}\cop{)}};
  \node[treenode]       (b) at (1,-1)  {\cfn{process\_slots}\cop{(}\cfield{state}\cop{,} \cfield{block.slot}\cop{)}};
  \node[treenode]       (c) at (2,-2)  {\cfn{process\_slot}\cop{(}\cfield{state}\cop{)}};
  \node[treenode]       (d) at (2,-3)  {\cfn{process\_epoch}\cop{(}\cfield{state}\cop{)}};
  \node[treenode]       (e) at (3,-4)  {\cfn{process\_justification\_and\_finalization}\cop{(}\cfield{state}\cop{)}};
  \node[treenode]       (f) at (3,-5)  {\cfn{process\_rewards\_and\_penalties}\cop{(}\cfield{state}\cop{)}};
  \node[treenode,gray]  (g) at (3,-6)  {\ldots\ 12 more epoch-level functions};
  \node[treenode]       (h) at (1,-7)  {\cfn{process\_block}\cop{(}\cfield{state}\cop{,} \cfield{block}\cop{)}};
  \node[treenode]       (i) at (2,-8)  {\cfn{process\_block\_header}\cop{(}\cfield{state}\cop{,} \cfield{block}\cop{)}};
  \node[treenode,gray]  (j) at (2,-9)  {\ldots\ 3 more block-level functions};
  \node[treenode]       (k) at (2,-10) {\cfn{process\_operations}\cop{(}\cfield{state}\cop{,} \cfield{block.body}\cop{)}};
  \node[treenode]       (l) at (3,-11) {\cfn{process\_proposer\_slashing}\cop{(}\ldots\cop{)}};
  \node[treenode,gray]  (m) at (3,-12) {\ldots\ 5 more operation handlers};
  \node[treenode]       (n) at (3,-13) {\cfn{process\_sync\_aggregate}\cop{(}\ldots\cop{)}};
  \draw[guide] (0.3,-0.45) -- (0.3,-7); \draw[guide] (0.3,-1) -- (0.9,-1); \draw[guide] (0.3,-7) -- (0.9,-7);
  \draw[guide] (1.3,-1.45) -- (1.3,-3); \draw[guide] (1.3,-2) -- (1.9,-2); \draw[guide] (1.3,-3) -- (1.9,-3);
  \draw[guide] (2.3,-3.45) -- (2.3,-6); \draw[guide] (2.3,-4) -- (2.9,-4); \draw[guide] (2.3,-5) -- (2.9,-5); \draw[guide] (2.3,-6) -- (2.9,-6);
  \draw[guide] (1.3,-7.45) -- (1.3,-10); \draw[guide] (1.3,-8) -- (1.9,-8); \draw[guide] (1.3,-9) -- (1.9,-9); \draw[guide] (1.3,-10) -- (1.9,-10);
  \draw[guide] (2.3,-10.45) -- (2.3,-13); \draw[guide] (2.3,-11) -- (2.9,-11); \draw[guide] (2.3,-12) -- (2.9,-12); \draw[guide] (2.3,-13) -- (2.9,-13);
\end{tikzpicture}
\par\medskip

\noindent First, \specfn{process\_slots} advances the state slot-by-slot from its current slot to the block's target slot,
applying per-slot (\specfn{process\_slot}) and per-epoch (\specfn{process\_epoch}) transitions at each step.
Second, \specfn{process\_block} applies block-specific updates through a sequence of sub-functions,
each validating one aspect of the block and updating the state.

The specification defines an additional \specplain{validate\_result} flag (default \specplain{True})
that controls whether the block signature and the post-state root are checked against the values the block carries.
Validators selected as proposers set this flag to \specplain{False} during block construction
to avoid a circular dependency~\cite{ethereumConsensusSpec},
whereas attesters use \specplain{True} when validating received blocks.

\subsection{Consensus Spectests}

To guard against cross-client discrepancies, the Ethereum Foundation~\cite{ethereumfoundation} maintains an official consensus test suite, \ctests{}~\cite{consensusSpecTests}.
Each test pairs an $(S, B)$ input with its expected outcome.
Developers write the inputs by hand, targeting specific scenarios
such as \specplain{proposer\_slashing},
while \cspec{} serves as an oracle to compute the expected output.
Coverage is therefore determined by the scenarios developers chose to write
rather than by any systematic analysis of the specification's validity conditions.

\subsection{Consensus Faults and Failures}

We distinguish a divergence's latent cause, the fault,
from its observable form, the failure.
Let $f^*$ denote the correct implementation of the state-transition function as defined by the specification.
A \textit{consensus fault} in a client $C$ is a deviation of $f_C$ from $f^*$:
\[
\exists\, (S, B) \text{ such that } f_C(S, B) \neq f^*(S,B)
\]
A fault is a latent property of the implementation,
independent of whether the triggering input is ever encountered in practice.

When a fault manifests as observable disagreement between clients on the same input,
we call it a \textit{consensus failure} (\textit{Class~A}).
If $f_{C_1}(S, B) \neq f_{C_2}(S, B)$,
then a consensus fault exists in one or both of $C_1$ and $C_2$.
A client may also terminate abnormally instead of returning a state or $\bot$,
which we likewise count as disagreement.
Identifying which client deviates from $f^*$ requires comparing outputs against the specification,
which we do in \cref{sec:eval}.

Every consensus failure implies a fault, but not every fault produces a consensus failure.
A fault that only manifests when \specplain{validate\_result=False}
causes a proposer to produce an invalid block.
Attesters running \specplain{validate\_result=True} reject the block,
but agree with each other.
This is a \textit{liveness failure} (\textit{Class~B}):
the proposer loses its block reward,
but no disagreement arises among attesters.

\subsection{\spectec}\label{sec:spectec}
\spectec~\cite{spectec,p4spectec} is a mechanization framework for language specifications.
It consists of a domain-specific language (DSL)
and a toolchain that generates interpreters, tests, and prose documents
from specifications written in the DSL.
\spectec was originally designed to mechanize the WebAssembly (Wasm) specification~\cite{spectec}, and P4-SpecTec is a redesign of \spectec centered on algorithmic inference rules~\cite{p4spectec}.
While this work uses P4-SpecTec, for the sake of brevity, this paper refers to it simply as \spectec.

Specifications written in the \spectec DSL are organized as a set of \textit{rules},
each defining one case of a relation.
The rule below illustrates the syntax:
  \begingroup
  \fvset{fontsize=\footnotesize, frame=lines, framesep=4pt,
         numbers=left, numbersep=5pt, xleftmargin=10pt}%
  \renewcommand{\theFancyVerbLine}{\tiny\color{gray}\arabic{FancyVerbLine}}%
  \begin{Verbatim}[commandchars=\\\{\}]
\PY{k}{rule}\PY{n+nf}{ }\PY{n+nf}{WeighJustification/justify\PYZus{}prev}\PY{p}{:}
  \PY{n}{state} \PY{n}{total\PYZus{}bal} \PY{n}{prev\PYZus{}bal} \PY{n}{cur\PYZus{}bal} \PY{o}{\PYZti{}\PYZgt{}} \PY{n}{state}\PY{p}{[}\PY{p}{.}\PY{n+na}{CUR\PYZus{}JCP} \PY{o}{=} \PY{n}{prev\PYZus{}epoch}\PY{p}{]}
  \PY{o}{\PYZhy{}\PYZhy{}} \PY{k}{if} \PY{n}{total\PYZus{}bal} \PY{o}{\PYZlt{}=} \PY{n+nf}{\PYZdl{}UINT64\PYZus{}MAX} \PY{o}{/} \PY{l+m+mi}{2}    \PY{c+c1}{;; inserted: overflow}
  \PY{o}{\PYZhy{}\PYZhy{}} \PY{k}{if} \PY{n}{prev\PYZus{}bal} \PY{o}{*} \PY{l+m+mi}{3} \PY{o}{\PYZgt{}=} \PY{n}{total\PYZus{}bal} \PY{o}{*} \PY{l+m+mi}{2}
  \PY{o}{\PYZhy{}\PYZhy{}} \PY{k}{if} \PY{o}{\PYZti{}}\PY{p}{(}\PY{n}{cur\PYZus{}bal} \PY{o}{*} \PY{l+m+mi}{3} \PY{o}{\PYZgt{}=} \PY{n}{total\PYZus{}bal} \PY{o}{*} \PY{l+m+mi}{2}\PY{p}{)}
  \PY{o}{\PYZhy{}\PYZhy{}} \PY{k}{if} \PY{n}{prev\PYZus{}epoch} \PY{o}{=} \PY{n+nf}{\PYZdl{}get\PYZus{}prev\PYZus{}epoch}\PY{p}{(}\PY{n}{state}\PY{p}{)}
\end{Verbatim}
  \endgroup

Line~1 is the rule name.
Line~2 is the \textit{conclusion}:
the relation instance being defined,
which updates the state's \specfield{CUR\_JCP} field.
Lines~3--6 are the \textit{premises}:
the conditions that must hold for the conclusion to apply.
However, not all premises represent validity conditions.
Line~6 binds \specplain{prev\_epoch} to the result of
the function call \specplain{\PY{n+nf}{\PYZdl{}get\PYZus{}prev\PYZus{}epoch}\PY{p}{(}\PY{n}{state}\PY{p}{)}}:
it introduces a new name for the result of a pure computation.
The other premises are boolean conditions.
If the evaluation of a boolean condition yields false, the transition becomes invalid.
We refer to bindings of purely computational results as \textit{let-premises},
and boolean conditions as \textit{if-premises}.
As we present in \cref{sec:coverage}, premise coverage is defined only over if-premises.
Finally, the comment on line~3 indicates that the premise has been inserted.
In other words, \cspec{} does not have a corresponding condition;
instead, it relies on an implicit exception when
\specplain{total\_bal * 2} overflows.
We discuss this class of premise insertion in \cref{sec:spec}.

\section{Overview}\label{sec:overview}


Fig.~\ref{fig:overview} illustrates an overview of \framework{}.
The foundation of \framework{} is \textbf{\mech{}}~(\cref{sec:spec}),
the mechanized specification of the Ethereum consensus algorithm
in the \spectec DSL.
It makes branch conditions and implicit exceptions explicit as if-premises.
From this, the \spectec toolchain generates an executable interpreter,
over which the remaining stages operate.

In the \textbf{Premise Coverage}~(\cref{sec:coverage}) stage,
we collect seed inputs from \ctests{},
selecting tests that target the \specfn{state\_transition} function,
and flatten multi-block sequences into individual $(S, B)$ pairs.
We then measure \emph{premise coverage}:
for each if-premise in the mechanized specification,
we record whether the seed suite evaluates it to true, to false, or to both.
We manually inspect the never-falsified premises
to exclude unfalsifiable conditions,
yielding the set of target premises.

In the \textbf{Test Generation} stage,
the generator propagates provenance from input fields to runtime values~(\cref{sec:provenance}),
negates the condition of each target premise to identify which fields to mutate~(\cref{sec:suggestion}),
and samples values from the resulting intervals~(\cref{sec:mutation}),
producing mutated $(S, B)$ pairs.

In the \textbf{Differential Testing}~(\cref{sec:diff_testing}) stage,
the generated tests are executed on six implementations,
the five consensus clients and \cspec{}.
Any disagreement among the clients is a divergence,
and \cspec{} serves as the reference for fault attribution.

\begin{figure}[t]
   \centering
   \begin{subfigure}{\linewidth}
  \begingroup
  \fvset{fontsize=\footnotesize, frame=lines, framesep=4pt,
         numbers=left, numbersep=5pt, xleftmargin=10pt}%
  \renewcommand{\theFancyVerbLine}{\tiny\color{gray}\arabic{FancyVerbLine}}%
  \input{listings/gen/weigh_py.tex}%
  \endgroup

      \caption{in~\cspec{}.}
      \label{fig:weigh-justification-and-finalization}
   \end{subfigure}

   \vspace{0.5em}
   \begin{subfigure}{\linewidth}
  \begingroup
  \fvset{fontsize=\footnotesize, frame=lines, framesep=4pt,
         numbers=left, numbersep=5pt, xleftmargin=10pt}%
  \renewcommand{\theFancyVerbLine}{\tiny\color{gray}\arabic{FancyVerbLine}}%
  \input{listings/gen/weigh_spectec.tex}%
  \endgroup

      \caption{in~\mech{}.}
      \label{fig:weigh-justification-and-finalization-spectec}
   \end{subfigure}
   \caption{\specfn{weigh\_justification\_and\_finalization} (simplified).}
   \label{fig:weigh-justification-and-finalization-all}
\end{figure}

\section{Mechanized Specification}\label{sec:spec}

\mech{} makes three aspects of \cspec{} explicit:
mutation semantics, branching structure, and exception boundaries.
We demonstrate these aspects with a simplified version of the function
\specfn{weigh\_justification\_and\_finalization}
in Fig.~\ref{fig:weigh-justification-and-finalization}.
For illustration purposes, we show only the core justification logic,
whereas the original \specfn{weigh\_justification\_and\_finalization} function
additionally handles finalization.

Lines 4--5 and 6--7 are independent if-blocks---not if-else---that
both update \specfield{cur\_justified\_cp}.
If both conditions hold,
the second silently overwrites the first.
Because such if-blocks interact through a shared state
but are not connected by control flow,
a branch-coverage tool can execute each block's true and false
branches, achieving 100\% coverage, without testing specific combinations
that determine which checkpoints are justified.
Also, the multiplication on lines 4 and 6
overflows when the total active balance exceeds $2^{63}$.
However, \cspec{} does not prevent such cases.

Fig.~\ref{fig:weigh-justification-and-finalization-spectec}
shows how \mech{} addresses these.

\PP{Mutation.}
\mech{} represents all state-modifying operations as relations defined by rules,
where each rule explicitly specifies the fields it modifies in its conclusion.
For example,
the conclusion of
\specfn{justify\_prev} (line 7)
shows that only \specfield{CUR\_JCP} is updated.
In contrast,
determining this within \cspec{}
requires tracing all possible branches.

\PP{Branches.}
\mech{} encodes each execution path as a separate rule.
For example, the relation \specfn{WeighJustification} consists of three rules,
each representing one reachable result of justification.
The premises of each rule (e.g., lines~3--5)
state the exact conditions under which this rule is taken,
eliminating the need to simulate overlapping if-blocks.
The three rules also make it clear that only three of four combinations in Fig.~\ref{fig:weigh-justification-and-finalization} are meaningful.

\PP{Exception Boundaries.}
\mech{} inserts overflow guards as explicit if-premises,
as shown on lines 3, 8, and 14.
Similarly,
bounds checks for array accesses are added
where \cspec{} relies on implicit exceptions.
We follow one such premise, the overflow check on line 8,
as a running example through premise coverage (\cref{sec:coverage})
and test generation (\cref{sec:testgen}).

Together, these three aspects
collectively provide the structural foundation
on which we define premise coverage and guide test generation.
In total,
\mech{} consists of \numLOC{} lines of code,
including \numRel{} state-modifying relations,
\numFunc{} helper functions,
and \numTotalIfPremises{} premises.
Of these premises, \numInsertedChecks{} are guards we insert
where \cspec{} would raise exceptions:
\numOverOrUnderflowChecks{} guard fixed-width arithmetic against overflow or underflow,
\numBoundsChecks{} guard index accesses or require two lists to agree in length,
and \numDivisorChecks{} requires a divisor to be non-zero.
We validate the correctness of \mech{} against \cspec{}
on \numValidationTests{} tests from the official test suite:
\numStateTransitionTests{} state-transition tests and \numUnitTests{} unit tests covering 22 specification functions.
We exclude tests targeting deprecated functions,
features outside the state transition (e.g., fork choice and light client),
and serialization.

\section{Premise Coverage}\label{sec:coverage}
For each if-premise,
premise coverage tracks whether \ctests{} evaluate it to true or false.
A premise is fully covered when it is evaluated to both values.
This metric is motivated by clause coverage~\cite{ammann2016,chilenski1994},
where each clause in a compound condition
must evaluate to both true and false.
In the \spectec DSL,
the compound condition of each rule
is the conjunction of its if-premises,
and each if-premise is an individual clause.

Evaluating a premise to false is particularly informative
because an if-premise can play one of two roles,
depending on the other rules in the relation.
If the premise appears identically across all rules,
such as the overflow checks on lines~3, 8, and~14
in Fig.~\ref{fig:weigh-justification-and-finalization-spectec},
it acts as an assertion:
failing it rejects the transition regardless of which rule is applied.
If the premise distinguishes one rule from another,
such as the justification condition on line~4,
it acts as a branch condition:
failing it causes a different rule to be applied (line~9).
We target both roles in our test generation.

To illustrate the assertion case, we use the running example shown in Fig.~\ref{fig:running-example}.
Across the baseline \ctests{}, premise coverage reveals that
the overflow check evaluates to true 21 times and to false 0 times.
Since the total balance has never caused an overflow, the check is marked as a generation target.
The remainder of the paper traces this running example through each stage of \framework{}.

\begin{figure}[t]
\small
\centering
\setlength{\tabcolsep}{3pt}
\renewcommand{\arraystretch}{1.0}
\begin{tabularx}{\columnwidth}{@{}l >{\raggedright\arraybackslash}X@{}}
\toprule
\textbf{Stage} & \textbf{Running example} \\
\midrule
\textbf{Premise} \,(\S\ref{sec:spec})
  & \specop{-{}-} \speckw{if} \specplain{total\_bal} \specop{<=} \specfn{\$UINT64\_MAX} \specop{/} \specnum{2} \\
\addlinespace
\textbf{Coverage} \,(\S\ref{sec:coverage})
  & true: $21$, \ false: $0$ \\
\addlinespace
\textbf{Provenance} \,(\S\ref{sec:provenance})
  & $\Pi(\text{\specplain{total\_bal}}) = \newline \{(\textsc{State},$
    $[\text{\specfield{validators}}, 0, \text{\specfield{effective\_balance}}]),\; \newline \ldots\}$ \\
\addlinespace
\textbf{Suggestion} \,(\S\ref{sec:suggestion})
  & $(\text{\specfield{state.validators[0].effective\_balance}},$ \newline
    $\text{\specplain{ToConst}}(>,\ \text{\specfn{\$UINT64\_MAX}\specplain{/2}}))$ \\
\addlinespace
\textbf{Mutation} \,(\S\ref{sec:mutation})
  & $\text{\specfield{state.validators[0].effective\_balance}} \in$ \newline
    $[\,\text{\specfn{\$UINT64\_MAX}\specplain{/2 + 1}},\ \max_\tau\,]$ \\
\addlinespace
\textbf{Outcome} \,(\S\ref{sec:eval})
  & $f_{C}(S,B) \neq f_{C'}(S,B)$ \ (Cases~\caseOverflow{} of Table~\ref{tab:rq1_result_merged_updated}) \\
\bottomrule
\end{tabularx}
\caption{Running example tracing an overflow check premise through the stages of \framework{}.}
\label{fig:running-example}
\end{figure}

Not all if-premises contribute to test generation.
We manually inspect the never-falsified premises and exclude two classes:
\begin{itemize}[labelsep=0.5em,leftmargin=1em]
\item An if-premise is \textit{unfalsifiable}
if it cannot evaluate to false under any reachable state.
The most common case is a \textit{tautology}:
a bounds check inside a loop that iterates over an indexed array,
a computed value constrained to a fixed range,
or a premise in a function only invoked under assumptions that guarantee it.\footnote{
  We also classify three premises that are beyond the scope of this paper
  as unfalsifiable:
  1) \texttt{\footnotesize\cfn{ee\_execute\_and\_verify\_payload}} is
  a mock function that unconditionally returns true,
  2) \texttt{\footnotesize validate\_result=False} always holds due to our test configuration,
  and 3) a whistleblower-index parameter~\cite{whistleblower} cannot be supplied under the current protocol.
}
Another common case is a \textit{closing branch}:
the final rule in a relation
may contain a premise whose negation is covered by preceding rules,
making it trivially satisfied when reached.

\item An if-premise is \textit{unattempted}
if no seed test evaluates it to true,
leaving our mutation-based generator with no seed to work from.
\end{itemize}

Of the \numTotalIfPremises{} if-premises in \mech{},
we exclude \numUnfalsifiablePremises{} unfalsifiable premises
(\numTautologyPremises{} tautologies and \numClosingBranchPremises{} closing branches),
leaving \numTotalFalsifiablePremises{} falsifiable premises as test generation targets.
Of these, \spell{\numUnattemptedPremises} are unattempted;
they remain in the denominator as a limitation of the current approach.
We count each premise independently, without deduplicating premises
that share a failure path.

\section{Test Generation and Differential Testing}\label{sec:testgen}

The goal of the test generator is to produce inputs
that falsify the target premises
identified by the coverage analysis of \cref{sec:coverage}.
Given a seed input $(S, B)$
that reaches a target premise $p$ and evaluates it to true,
the generator determines which fields of $S$ or $B$ to mutate
and what values to substitute so that $p$ evaluates to false.
However, input fields and premise operands are connected only indirectly:
values pass through helper functions, arithmetic operations, and record constructions
before reaching a premise.

To address this, the generator operates in three phases.
Provenance propagation (\cref{sec:provenance})
tracks which input fields contribute to each runtime value,
suggestion extraction (\cref{sec:suggestion})
negates each target premise's constraint and pairs it with those fields,
and mutation application (\cref{sec:mutation})
turns each suggestion into concrete substitutions.
Differential testing (\cref{sec:diff_testing}) then executes the generated
tests across the clients and compares their outcomes.

\begin{figure}[t]
\vspace*{-1.4em}
\begin{algorithm}[H]
\small
\caption{Provenance Propagation}
\label{alg:provenance}
\begin{algorithmic}[1]
\Require Input $I = (S, B)$
\Ensure Every runtime value carries a provenance set $\Pi$
\State $\Pi(S) \gets \{(\textsc{State}, \epsilon)\}$;\; $\Pi(B) \gets \{(\textsc{Block}, \epsilon)\}$
\Statex \Comment{Propagation rules applied during interpretation:}
\State $\Pi(v.f) \gets \{(s, p \cdot f) \mid (s, p) \in \Pi(v)\}$ \Comment{field access}
\State $\Pi(v[i]) \gets \{(s, p \cdot i) \mid (s, p) \in \Pi(v)\}$ \Comment{index access}
\State $\Pi(v_1 \oplus v_2) \gets \Pi(v_1) \cup \Pi(v_2)$ \Comment{arithmetic}
\State $\Pi(r[f_i:v_i]) \gets \{(s, p \cdot f_i) \mid (s,p) \in \Pi(r)\}$ \Comment{record update}
\State $\Pi(f(v_1, \ldots, v_n)) \gets \bigcup_i \Pi(v_i)$ \; if $\Pi(f(v_1, \ldots, v_n)) = \emptyset$ \Comment{call}
\State $\Pi(\textit{literal}) \gets \emptyset$ \Comment{constants}
\end{algorithmic}
\end{algorithm}
\vspace*{-1em}
\end{figure}

\subsection{Provenance Propagation}\label{sec:provenance}

A \emph{provenance annotation} records how a runtime value
is derived from a JSON input.
Each annotation is a set of pairs $(\mathit{source}, \mathit{path})$,
where $\mathit{source} \in \{\textsc{State}, \textsc{Block}\}$ identifies the input root
and $\mathit{path}$ is a sequence of field-access and index-access steps.
For example, evaluating \specfield{state.validators[0].effective\_balance}
produces a value annotated with the provenance
$\{(\textsc{State},$
$[\specfield{validators}, 0, \specfield{effective\_balance}])\}$.
The running example's operand \specplain{total\_bal}
sums the active validators' \specfield{effective\_balance} fields,
so its provenance merges their paths (Fig.~\ref{fig:running-example}).

Algorithm~\ref{alg:provenance} summarizes the propagation rules,
which apply to every interpreter operation.
A value computed from several inputs retains all paths,
which ensures no data dependencies are missed.
Field access, indexing, and record update each extend a path by one step,
and arithmetic merges the annotations of its operands.
For function calls,
operations within the function body follow the same propagation rules,
so the return value is already annotated.
Since calls to built-in functions whose internals are opaque (e.g., hashing and signing) do not return provenance,
the algorithm falls back to merging the annotations of their arguments.

\subsection{Suggestion Extraction}\label{sec:suggestion}

The instrumented interpreter examines each target premise during execution
and produces a set of \emph{mutation suggestions}.
Each suggestion is a pair $(\mathit{path},\; \mathit{strategy})$
that specifies an input field and a negated constraint on its value.

\begin{figure}[t]
\vspace*{-1em}
\begin{algorithm}[H]
\small
\caption{Mutation Suggestion Extraction}
\label{alg:suggestion}
\begin{algorithmic}[1]
\Require Target premises $P_{\mathit{target}}$, instrumented execution trace
\Ensure Mutation suggestions $\mathcal{M}$
\State $\mathcal{M} \gets \emptyset$
\For{each if-premise $p \in P_{\mathit{target}}$ reached during execution}
    \State $\mathit{cond} \gets$ condition of $p$
    \If{$\mathit{cond}$ is $e_1 \prec e_2$} \Comment{comparison}
        \State $(v_1, \Pi_1) \gets \textsc{Eval}(e_1)$;\;
               $(v_2, \Pi_2) \gets \textsc{Eval}(e_2)$
        \For{$\pi \in \Pi_1$}
            $\mathcal{M} \gets \mathcal{M} \cup \{(\pi,\; \texttt{ToConst}(\overline{\prec},\; v_2))\}$
        \EndFor
        \For{$\pi \in \Pi_2$}
            $\mathcal{M} \gets \mathcal{M} \cup \{(\pi,\; \texttt{ToConst}(\overline{\succ},\; v_1))\}$
        \EndFor
    \ElsIf{$\mathit{cond}$ is $e_1 \wedge e_2$} \Comment{conjunction}
        \State recurse on $e_1$ and $e_2$ independently
    \ElsIf{$\mathit{cond}$ is $|e| \prec n$} \Comment{length comparison}
        \For{$\pi \in \Pi(e)$}
            $\mathcal{M} \gets \mathcal{M} \cup \{(\pi,\; \texttt{ToLength}(\overline{\prec},\; n))\}$
        \EndFor
    \ElsIf{$\mathit{cond}$ is $\textit{call}(a_1, \ldots, a_n)$} \Comment{opaque call}
        \For{$a_j$ with $\Pi(a_j) \neq \emptyset$}
            \For{$\pi \in \Pi(a_j)$}
                $\mathcal{M} \gets \mathcal{M} \cup \{(\pi,\; \texttt{Unknown})\}$
            \EndFor
        \EndFor
    \EndIf
\EndFor
\State \Return $\mathcal{M}$
\end{algorithmic}
\end{algorithm}
\end{figure}

Algorithm~\ref{alg:suggestion} summarizes the procedure,
which dispatches on the condition of each premise.
For a comparison $e_1 \prec e_2$,
it evaluates both sides to concrete values $v_1$ and $v_2$
with provenance sets $\Pi_1$ and $\Pi_2$.
Then, for each $\pi \in \Pi_1$, it emits
$(\pi,\; \specplain{ToConst}(\overline{\prec},\; v_2))$,
where $\overline{\prec}$ negates $\prec$ (e.g., $\overline{<} = {\geq}$),
and symmetrically for each $\pi \in \Pi_2$.
%
In the running example, negating the overflow check's \specplain{<=} to \specplain{>}
yields a suggestion to raise a representative entry,
\specfield{validators[0].effective\_balance}, above \specfn{\$UINT64\_MAX}\specplain{/2}
(Fig.~\ref{fig:running-example}).
%
Conjunctions ($e_1 \wedge e_2$) are decomposed:
each conjunct is processed independently,
since falsifying either one falsifies the conjunction.
Negations and boolean-equality patterns
(e.g., $\neg(x = \mathit{false})$) are simplified before extraction.
Length expressions ($|e| \prec n$)
produce \specplain{ToLength} suggestions
that target the list's size rather than its element values.
Function calls with opaque semantics
(e.g., \specfn{bls\_verify})
produce \specplain{Unknown} suggestions for each argument path,
resolved during mutation application
by generating type-compatible random values.

\subsection{Mutation Application}\label{sec:mutation}

The mutation application stage translates the suggestion set $\mathcal{M}$
from Algorithm~\ref{alg:suggestion} into concrete test inputs $\mathcal{T}$.
Multiple suggestions often target the same input field,
because a single field may appear in several premises
or on both sides of a comparison.

Algorithm~\ref{alg:apply} gives the full procedure.
A type tree $\Gamma$, pre-computed by a static pass over \mech{},
maps each field path to its type, extracted from the relation
and function signatures in the DSL.
Suggestions are grouped by \emph{structural key},
field paths with all array indices normalized.
This prevents per-element duplication
when a premise references a field inside a loop
(e.g., \specfield{validators[i].effective\_balance} for each index $i$).
Within each structural-key group,
suggestions are further partitioned by kind:
\specplain{ToConst} and \specplain{ToLength} suggestions
form separate subgroups,
while \specplain{Unknown} suggestions contribute
slice points to both.
For each group,
it collects the \specplain{ToConst} and \specplain{ToLength} constraints into $\mathit{known}$
and the \specplain{Unknown} values into $\mathit{unknown}$, respectively (lines 3--4).

It computes base intervals $\mathcal{R}$ (line~6)
by mapping each $(\mathit{op}, n)$ to one or two intervals within
the field's representable range $[\min_\tau, \max_\tau]$.
For example,
$(\geq,5)$ yields $[5, \max_\tau]$,
and $(\neq,30)$ yields $[0, 29] \cup [31, \max_\tau]$.
All intervals from the same subgroup are merged
by sorting and unioning overlapping or adjacent ranges.

It then subdivides them into final intervals $\mathcal{F}$ (line~9).
An \specplain{Unknown} value is a value used at runtime,
inside a call or expression the extraction fails to inspect.
$\textsc{Slice}$ therefore treats each $u \in \mathit{unknown}$ as a \emph{slice point},
inserting $\{u{-}1, u, u{+}1\}$ into the set of interval boundaries.
The merged boundaries then form the final intervals:
consecutive pairs $[b_i, b_{i+1}]$ contained within some base interval.

\begin{figure}[t]
\vspace*{-1em}
\begin{algorithm}[H]
\small
\caption{Mutation Application}
\label{alg:apply}
\begin{algorithmic}[1]
\Require Suggestions $\mathcal{M}$, seed input $I = (S, B)$, type tree $\Gamma$
\Ensure Mutated test inputs $\mathcal{T}$
\State $\mathcal{T} \gets \emptyset$
\For{each field path $\pi$ and kind $\mathit{kind} \in \{\texttt{ToConst}, \texttt{ToLength}\}$}
    \State $\mathit{known} \gets \{(\mathit{op}, n) \mid (\pi, \mathit{kind}(\mathit{op}, n)) \in \mathcal{M}\}$
    \State $\mathit{unknown} \gets \{v \mid (\pi, \texttt{Unknown}(v)) \in \mathcal{M}\}$
    \If{$\mathit{known} = \emptyset$ and $\mathit{unknown} = \emptyset$}
        \textbf{continue}
    \EndIf
    \State $\mathcal{R} \gets \textsc{Merge}\bigl(\bigcup_{(\mathit{op},\, n) \in \mathit{known}} \textsc{Intervals}(\mathit{op},\; n,\; \min_\tau,\; \max_\tau)\bigr)$
    \If{$\mathit{known} \neq \emptyset$ and $\mathcal{R} = \emptyset$}
        \textbf{continue}
    \EndIf
    \If{$\mathcal{R} = \emptyset$}
        $\mathcal{R} \gets \{[\min_\tau,\; \max_\tau]\}$
    \EndIf
    \State $\mathcal{F} \gets \textsc{Slice}(\mathcal{R},\; \mathit{unknown})$
    \State $V \gets \textsc{Sample}(\mathcal{F})$
    \State $V \gets \{v \in V \mid \textsc{TypeOk}(v,\; \Gamma[\pi]) \wedge v \neq I[\pi]\}$
    \If{$V = \emptyset$}
        $V \gets \textsc{Fallback}(I[\pi],\; \Gamma[\pi])$
    \EndIf
    \For{$v \in V$}
        $\mathcal{T} \gets \mathcal{T} \cup \{I[\pi \mapsto v]\}$
    \EndFor
\EndFor
\State \Return $\mathcal{T}$
\end{algorithmic}
\end{algorithm}
\vspace*{-1em}
\end{figure}

It samples concrete substitution values $V$ from $\mathcal{F}$ (line~10)
using Algorithm~\ref{alg:sample},
which samples three classes of values.
Boundary values (line~3, Algorithm~\ref{alg:sample}) are the endpoints of each final interval,
testing the exact points where the premise makes a transition between true and false.
Transition values (lines~4--5) are the nearest values outside the final intervals,
confirming that the original constraint holds at the complement.
Interior values (lines~6--8) are evenly spaced within intervals
that exceed the given minimum width.
Then, it discards any transition value that falls inside another interval.

Each sampled value is checked against the field's JSON type before substitution
(line 11, Algorithm~\ref{alg:apply}).
A mismatch can occur because provenance tracking is over-approximated
for compound expressions and function calls.
If all values for a given path are incompatible,
a fallback produces type-appropriate mutations (line~12),
such as toggling booleans,
incrementing or decrementing integers,
or appending to or truncating lists.

\begin{figure}[t]
\vspace*{-1em}
\begin{algorithm}[H]
\small
\caption{Interval Sampling}
\label{alg:sample}
\begin{algorithmic}[1]
\Require Final intervals $\mathcal{F} = \{[\ell_1, h_1], \ldots, [\ell_m, h_m]\}$, minimum interior width $w$
\Ensure Sample set $V$
\State $V \gets \emptyset$;\; $c \gets$ \textbf{if} $|\mathcal{F}| = 1$ \textbf{then} $2$ \textbf{else} $1$
\For{$[\ell_i, h_i] \in \mathcal{F}$}
    \State $V \gets V \cup \{\ell_i,\; h_i\}$ \Comment{boundary values}
    \If{$\ell_i - 1 \geq \min_\tau$}
        $V \gets V \cup \{\ell_i - 1\}$ \Comment{transition values}
    \EndIf
    \If{$h_i + 1 \leq \max_\tau$}
        $V \gets V \cup \{h_i + 1\}$
    \EndIf
    \If{$h_i - \ell_i \geq w$} \Comment{interior values}
        \For{$j = 1, \ldots, c$}
            \State $V \gets V \cup \bigl\{\ell_i + \lfloor (h_i - \ell_i) \cdot j\, /\, (c+1) \rfloor \bigr\}$
        \EndFor
    \EndIf
\EndFor
\State $V \gets V \setminus \{v \mid v \in [\ell_j,\; h_j] \text{ for some } [\ell_j, h_j] \in \mathcal{F}\}$
\Statex \Comment{remove transition values that land inside an interval}
\State \Return $V$

\end{algorithmic}
\end{algorithm}
\vspace*{-1em}
\end{figure}

\begin{table*}[t]
  \centering
  \scriptsize
  \captionsetup{width=\textwidth}
  \begin{threeparttable}
  \caption{Cross-client divergence cases exposed by \framework{}.}
  \label{tab:rq1_result_merged_updated}
  \setlength{\tabcolsep}{2.5pt}
  \setlength{\arrayrulewidth}{0.5pt}
  \renewcommand{\arraystretch}{1.5}
  \renewcommand{\tabularxcolumn}[1]{m{#1}}
  \newcommand{\resthree}[1]{%
    \implres{#1}{$\checkmark$}{\scalebox{0.45}{$\bullet\!\bullet\!\bullet$}}%
  }
  \begin{tabularx}{\textwidth}{
    @{}
    >{\RaggedRight\arraybackslash}m{0.18\textwidth}
    >{\centering\arraybackslash}m{0.035\textwidth}
    >{\RaggedRight\arraybackslash}X
    >{\centering\arraybackslash}m{0.15\textwidth}
    >{\centering\arraybackslash}m{0.15\textwidth}
    >{\centering\arraybackslash}m{0.045\textwidth}
    @{}
  }
  \hline
  \textbf{Root-Cause Category} & \textbf{Case} & \textbf{Description} & \textbf{\texttt{validate\_result=False}} & \textbf{\texttt{validate\_result=True}} & \textbf{Class} \\
  \hline

  \multirow{3}{=}{Precondition Violation}
    & 1
    & \texttt{\cfield{state.validators}} set to an empty list.
    & \mbox{[\resfail{C}\,\resfail{L}\,\resfail{Lo}\,\rescrash{N}\,\resfail{P}\,\resfail{T}]}
    & \mbox{[\resfail{C}\,\resfail{L}\,\resfail{Lo}\,\rescrash{N}\,\resfail{P}\,\resfail{T}]}
    & A \\ 
    & 2
    & \texttt{\cfield{state.slot}} set equal to \texttt{\cfield{block.slot}}.
    & \mbox{[\resfail{C}\,\ressuc{L}\,\ressuc{Lo}\,\resfail{N}\,\ressuc{P}\,\resfail{T}]}
    & \mbox{[\resfail{C}\,\resfail{L}\,\resfail{Lo}\,\resfail{N}\,\resfail{P}\,\resfail{T}]}
    & B \\
    & 3
    & The deposit index set above the deposit count by changing either field.
    & \mbox{[\resfail{C}\,\ressuc{L}\,\resfail{Lo}\,\resfail{N}\,\resfail{P}\,\resfail{T}]}
    & \mbox{[\resfail{C}\,\resfail{L}\,\resfail{Lo}\,\resfail{N}\,\resfail{P}\,\resfail{T}]}
    & B \\
  \hline

  \multirow{4}{=}{Lossy Decoding}
    & 4
    & Each of eight zero-valued state \texttt{uint64} fields set to its max/near-max value. 
    & \mbox{[\resfail{C}\,\resfail{L}\,\ressuc{Lo}\,\resfail{N}\,\resfail{P}\,\resfail{T}]}
    & \mbox{[\resfail{C}\,\resfail{L}\,\ressuc{Lo}\,\resfail{N}\,\resfail{P}\,\resfail{T}]}
    & A \\
    & 5
    & Each of two zero-valued state \texttt{uint64} fields set to its max/near-max value.
    & \mbox{[\resone{C}\,\resone{L}\,\restwo{Lo}\,\resone{N}\,\resone{P}\,\resone{T}]}
    & \mbox{[\resfail{C}\,\resfail{L}\,\ressuc{Lo}\,\resfail{N}\,\resfail{P}\,\resfail{T}]}
    & A \\
    & 6
    & A balance entry or validator \texttt{uint64} field set to a large value.
    & \mbox{[\resone{C}\,\resone{L}\,\restwo{Lo}\,\resone{N}\,\resone{P}\,\resone{T}]}
    & \mbox{[\resfail{C}\,\resfail{L}\,\resfail{Lo}\,\resfail{N}\,\resfail{P}\,\resfail{T}]}
    & B \\
    & 7
    & \texttt{\cfield{state.slot}} set to the max \texttt{uint64} value.
    & \mbox{[\resfail{C}\,\resfail{L}\,\ressuc{Lo}\,\resfail{N}\,\resfail{P}\,\resfail{T}]}
    & \mbox{[\resfail{C}\,\resfail{L}\,\resfail{Lo}\,\resfail{N}\,\resfail{P}\,\resfail{T}]}
    & B \\[0.4ex]
  \hline

  Configuration Inconsistency
    & 8
    & \texttt{\cfield{state.fork.current\_version}} is rolled back to pre-Capella era.
    & \mbox{[\resfail{C}\,\resfail{L}\,\ressuc{Lo}\,\resfail{N}\,\resfail{P}\,\resfail{T}]}
    & \mbox{[\resfail{C}\,\resfail{L}\,\resfail{Lo}\,\resfail{N}\,\resfail{P}\,\resfail{T}]}
    & B \\
  \hline

  \multirow{7}{=}{Numeric Underflow/Overflow}
    & 9
    & \texttt{\cfield{finalized\_checkpoint.epoch}} set to the max \texttt{uint64} value.
    & \mbox{[\resfail{C}\,\resfail{L}\,\restwo{Lo}\,\resone{N}\,\resone{P}\,\resfail{T}]}
    & \mbox{[\resfail{C}\,\resfail{L}\,\ressuc{Lo}\,\resfail{N}\,\resfail{P}\,\resfail{T}]}
    & A \\
    & 10
    & \texttt{\cfield{finalized\_checkpoint.epoch}} set to exceed \texttt{previous\_epoch}.
    & \mbox{[\resfail{C}\,\resfail{L}\,\restwo{Lo}\,\resone{N}\,\resone{P}\,\resfail{T}]}
    & \mbox{[\resfail{C}\,\resfail{L}\,\resfail{Lo}\,\resfail{N}\,\resfail{P}\,\resfail{T}]}
    & B \\
    & 11
    & A \texttt{\cfield{state.balances}} entry set to the max \texttt{uint64} value.
    & \mbox{[\resfail{C}\,\resfail{L}\,\restwo{Lo}\,\resone{N}\,\resone{P}\,\resfail{T}]}
    & \mbox{[\resfail{C}\,\resfail{L}\,\resfail{Lo}\,\resfail{N}\,\resfail{P}\,\resfail{T}]}
    & B \\
    & 12
    & A \texttt{\cfield{state.inactivity\_scores}} entry set to the max \texttt{uint64} value.
    & \mbox{[\resfail{C}\,\resfail{L}\,\restwo{Lo}\,\resone{N}\,\resfail{P}\,\resfail{T}]}
    & \mbox{[\resfail{C}\,\resfail{L}\,\resfail{Lo}\,\resfail{N}\,\resfail{P}\,\resfail{T}]}
    & B \\
    & 13
    & A \texttt{\cfield{state.balances}} entry set to a near-max \texttt{uint64} value.
    & \mbox{[\resfail{C}\,\resfail{L}\,\ressuc{Lo}\,\ressuc{N}\,\resfail{P}\,\resfail{T}]}
    & \mbox{[\resfail{C}\,\resfail{L}\,\resfail{Lo}\,\resfail{N}\,\resfail{P}\,\resfail{T}]}
    & B \\
    & 14
    & A \texttt{\cfield{state.balances}} entry set to a near-max \texttt{uint64} value.
    & \mbox{[\resfail{C}\,\resfail{L}\,\restwo{Lo}\,\resone{N}\,\resfail{P}\,\resfail{T}]}
    & \mbox{[\resfail{C}\,\resfail{L}\,\resfail{Lo}\,\resfail{N}\,\resfail{P}\,\resfail{T}]}
    & B \\
    & 15
    & A \texttt{\cfield{state.slashings}} entry set to a large \texttt{uint64} value.
    & \mbox{[\resfail{C}\,\resfail{L}\,\ressuc{Lo}\,\ressuc{N}\,\ressuc{P}\,\resfail{T}]}
    & \mbox{[\resfail{C}\,\resfail{L}\,\resfail{Lo}\,\resfail{N}\,\resfail{P}\,\resfail{T}]}
    & B \\
  \hline

  State-Update Inconsistency
    & 16
    & \texttt{\cfield{state.justification\_bits[0]}} flipped from \texttt{false} to \texttt{true}.
    & \mbox{[\resone{C}\,\resone{L}\,\resone{Lo}\,\resone{N}\,\restwo{P}\,\resone{T}]}
    & \mbox{[\resfail{C}\,\resfail{L}\,\resfail{Lo}\,\resfail{N}\,\resfail{P}\,\resfail{T}]}
    & B \\
  \hline

  \multirow{11}{=}{Invariant Enforcement Inconsistency}
    & 17
    & A validator \texttt{\cfield{effective\_balance}} set to the max \texttt{uint64} value.
    & \mbox{[\resfail{C}\,\resfail{L}\,\restwo{Lo}\,\resone{N}\,\resone{P}\,\resfail{T}]}
    & \mbox{[\resfail{C}\,\resfail{L}\,\resfail{Lo}\,\resfail{N}\,\resfail{P}\,\resfail{T}]}
    & B \\
    & 18
    & A validator \texttt{\cfield{effective\_balance}} set to a large non-canonical value.
    & \mbox{[\resfail{C}\,\resone{L}\,\restwo{Lo}\,\resfail{N}\,\resfail{P}\,\resfail{T}]}
    & \mbox{[\resfail{C}\,\resfail{L}\,\resfail{Lo}\,\resfail{N}\,\resfail{P}\,\resfail{T}]}
    & B \\
    & 19
    & A validator \texttt{\cfield{effective\_balance}} set to the max \texttt{uint64} value.
    & \mbox{[\resfail{C}\,\resfail{L}\,\restwo{Lo}\,\resone{N}\,\resfail{P}\,\resfail{T}]}
    & \mbox{[\resfail{C}\,\resfail{L}\,\resfail{Lo}\,\resfail{N}\,\resfail{P}\,\resfail{T}]}
    & B \\
    & 20
    & A validator \texttt{\cfield{effective\_balance}} set to a large non-canonical value.
    & \mbox{[\resone{C}\,\resfail{L}\,\restwo{Lo}\,\resone{N}\,\resone{P}\,\resone{T}]}
    & \mbox{[\resfail{C}\,\resfail{L}\,\resfail{Lo}\,\resfail{N}\,\resfail{P}\,\resfail{T}]}
    & B \\
    & 21
    & A validator \texttt{\cfield{effective\_balance}} set to a large non-canonical value.
    & \mbox{[\resfail{C}\,\resone{L}\,\restwo{Lo}\,\resone{N}\,\resone{P}\,\resfail{T}]}
    & \mbox{[\resfail{C}\,\resfail{L}\,\resfail{Lo}\,\resfail{N}\,\resfail{P}\,\resfail{T}]}
    & B \\
    & 22
    & A validator \texttt{\cfield{effective\_balance}} set to a non-canonical value.
    & \mbox{[\resone{C}\,\resone{L}\,\restwo{Lo}\,\resone{N}\,\resone{P}\,\resone{T}]}
    & \mbox{[\resfail{C}\,\resfail{L}\,\resfail{Lo}\,\resfail{N}\,\resfail{P}\,\resfail{T}]}
    & B \\
    & 23
    & \texttt{\cfield{state.balances}} set to an empty list while validators remain populated.
    & \mbox{[\resfail{C}\,\resfail{L}\,\ressuc{Lo}\,\resfail{N}\,\resfail{P}\,\resfail{T}]}
    & \mbox{[\resfail{C}\,\resfail{L}\,\resfail{Lo}\,\resfail{N}\,\resfail{P}\,\resfail{T}]}
    & B \\
    & 24
    & \texttt{\cfield{current\_epoch\_participation}} set to an empty list.
    & \mbox{[\resfail{C}\,\resfail{L}\,\restwo{Lo}\,\resfail{N}\,\resone{P}\,\resfail{T}]}
    & \mbox{[\resfail{C}\,\resfail{L}\,\resfail{Lo}\,\resfail{N}\,\resfail{P}\,\resfail{T}]}
    & B \\
    & 25
    & \texttt{\cfield{current\_epoch\_participation}} truncated to 63 entries.
    & \mbox{[\resfail{C}\,\resone{L}\,\restwo{Lo}\,\resfail{N}\,\resthree{P}\,\resfail{T}]}
    & \mbox{[\resfail{C}\,\resfail{L}\,\resfail{Lo}\,\resfail{N}\,\resfail{P}\,\resfail{T}]}
    & B \\
    & 26
    & \texttt{\cfield{previous\_epoch\_participation}} set to an empty list.
    & \mbox{[\resfail{C}\,\resfail{L}\,\ressuc{Lo}\,\resfail{N}\,\ressuc{P}\,\resfail{T}]}
    & \mbox{[\resfail{C}\,\resfail{L}\,\resfail{Lo}\,\resfail{N}\,\resfail{P}\,\resfail{T}]}
    & B \\
    & 27
    & \texttt{\cfield{previous\_epoch\_participation}} truncated to 63 entries.
    & \mbox{[\resfail{C}\,\restwo{L}\,\resone{Lo}\,\resfail{N}\,\resone{P}\,\resfail{T}]}
    & \mbox{[\resfail{C}\,\resfail{L}\,\resfail{Lo}\,\resfail{N}\,\resfail{P}\,\resfail{T}]}
    & B \\
  \hline

  \end{tabularx}

  \begin{tablenotes}[flushleft]
    \scriptsize
    \item \textbf{Implementations:} C (\cspec{}), L (Lighthouse), Lo (Lodestar), N (Nimbus), P (Prysm), and T (Teku).
    \item \textbf{Outcomes:} \suc(\textsc{Success}),
\fal(\textsc{Fail}), and
\cra(\textsc{Crash}).
    \item \textbf{Post-state markers:} Identical dot markers denote the same post-state group; different markers denote different groups.
    \item \textbf{Classification:} Class~A: consensus failure---discrepancy persists under full validation; Class~B: liveness failure---clients agree under full validation (\cref{sec:background})
  \end{tablenotes}
  \end{threeparttable}
\end{table*}

For the running example shown in Fig.~\ref{fig:running-example}, the suggestion on the \specplain{uint64} field
\specfield{validators[0].effective\_balance}
yields the base interval $[\specfn{\$UINT64\_MAX}\specplain{/2 + 1},\; \max_\tau]$.
Following Algorithm~\ref{alg:sample}, \framework{} samples
the two boundary values $\specfn{\$UINT64\_MAX}\specplain{/2 + 1}$ and $\max_\tau$,
the transition value $\specfn{\$UINT64\_MAX}\specplain{/2}$ just below the boundary,
and two interior values at the interval's thirds.
Mutating the field to these values produces tests that overflow
\specplain{total\_bal} and expose cross-client divergences.

Our test generator greedily selects \numSeeds{} seeds
among \numStateTransitionTests{} state-transition inputs from \ctests{}
to cover \numTargetPremises{} target premises, each up to seven times.
A total of \numGeneratedTests{} tests are generated.


\subsection{Differential Testing}\label{sec:diff_testing}

We execute the generated test cases on the five consensus clients
and \cspec{} under identical conditions.
Each client $C$ processes the same input
and returns either an updated state $S_C$ or $\bot_C$,
or terminates abnormally.
A divergence occurs when clients
produce different outcomes on the same input, including
(i) accept/reject disagreement,
(ii) inconsistent post-states ($S_{C_1} \neq S_{C_2}$), or
(iii) abnormal termination.
Because \mech{} is not used as an oracle, a specification error in it
does not lead to false positives in the discovered divergences.


\section{Evaluation}\label{sec:eval}
We evaluate \framework{} using the following research questions:

\begin{itemize}
   \setlength{\itemsep}{0.2em}
 \item [\textbf{RQ1}.]
   \textbf{Bug-Finding Effectiveness.} Can \framework{} uncover cross-client divergence cases?
 \item [\textbf{RQ2}.]
   \textbf{Diagnostic Power of Premise Coverage.} What does premise coverage reveal that code coverage metrics do not?
 \item [\textbf{RQ3}.]
   \textbf{Contribution of Specification Guidance.} Does \framework{} outperform unguided mutation under an equal budget?
 \item [\textbf{RQ4}.]
   \textbf{Cross-Fork Bug Reproducibility and Maintainability.} Do bugs in shared logic affect multiple forks?
\end{itemize}

The consensus specification defines state-transition functions
for seven fork versions (Phase0 through Fulu), where each fork incrementally 
extends the previous one. 
We evaluate on Capella throughout, and reproduce on Deneb in \cref{sec:rq4}.
Following the Fusaka mainnet announcement~\cite{Fusaka-Mainnet-Announcement},
we target the Ethereum consensus specification v1.6.0 and corresponding
mainnet-ready client versions:
Lighthouse v8.0.1,
Lodestar (state-transition) v1.36.0,
Nimbus v25.11.1,
Prysm v7.0.0,
and Teku v25.11.1.
All experiments are conducted on an Ubuntu machine with AMD Ryzen\textsuperscript{TM} 9 9950X CPU and 128\,GB RAM.

\subsection{RQ1: Bug-Finding Effectiveness}\label{sec:rq1}
Using \numGeneratedTests{} mutated test inputs generated by \framework{},
we identify \numDivergences{} cross-client divergences, 
as summarized in Table~\ref{tab:rq1_result_merged_updated}.
%
Inputs are grouped into a single case when they share the same 
implementation fault and all six implementations exhibit identical behaviors.
Similar inputs may be reported as different cases when they trigger different client behaviors.
%
Each case is first executed with \specplain{validate\_result=False} to expose the divergence,
then with \specplain{validate\_result=True} to classify it as Class~A or~B
(\cref{sec:background}).

\framework{} found \numClassA{} Class-A and \numClassB{} Class-B divergences,
all reported to the Ethereum Protocol Bug Bounty Program.

\PP{Precondition Violation}
In three cases, a client mishandles or fails to check
a condition that must hold
before invoking a function in the state transition.
In Case~1, the input state has no validators.
The other implementations reject it before selecting a committee,
but Nimbus continues into sync-committee selection
(\specfn{get\_next\_sync\_committee\_keys}),
which performs a modulo operation with the validator count
and terminates with a \code{SIGFPE}.
In Case~2, the pre-state slot already equals the block slot.
\specfn{process\_slots} requires it to be strictly smaller
and raises an error otherwise,
but Lighthouse, Prysm, and Lodestar apply the condition
only as the bound of their slot-advancement loop.
An equal slot therefore performs zero iterations rather than raising an error,
and the block is processed against a state
that never went through slot processing.
In Case~3, the input state has more processed deposits than deposits that exist
(\specfield{eth1\_deposit\_index} above \specfield{eth1\_data.deposit\_count}).
While \cspec{} rejects this immediately due to an underflow,
Lighthouse accepts such an input when the block carries zero deposits.

\PP{Lossy Decoding}
In four cases,
converting a serialized SSZ value into a client's internal representation
changes a value that should have been preserved.
All four originate in Lodestar's storage of these fields
as JavaScript numbers, but diverge by different mechanisms.
A JavaScript number cannot represent every 64-bit integer,
and the largest one is read back as zero.
Lodestar therefore runs the transition
on a different state than the one it was given.
In Case~4, the affected fields were zero before we mutated them,
so Lodestar reads back exactly the value we replaced
and processes the input state as if it were unmutated.
In Case~5, the mutated fields do not take part in the block's header checks,
so no implementation rejects the input
and the zeroing shows up only in the state Lodestar produces.
In Case~6, the value is large but short of the maximum,
so Lodestar rounds it to a nearby value rather than to zero.
In Case~7, Lodestar reads \specfield{state.slot} as zero
and advances the state through 33 slots,
whereas the other implementations reject the input.

\PP{Configuration Inconsistency}
In one case, a client selects the fork
from its local configuration
rather than the input state.
In Case~8, the input state names an earlier fork version
(\specfield{state.fork.current\_version}).
Signature checks derive their domain from this field
(\specfn{get\_domain}),
so under that version the block's signature should not verify.
Lodestar takes the version from its own configuration rather than the state,
so it fails to reject the invalid signature.

\PP{Numeric Underflow/Overflow}
In seven cases, an arithmetic operation produces a result
outside the 64-bit range,
and clients either do not check for this condition or handle it differently.
In Cases~9 and~10,
the distance between the finalized and previous epoch
is negative (\specfn{get\_finality\_delay}).
Prysm and Nimbus compute it with wrap-around arithmetic,
and Lodestar computes it as a JavaScript number,
so all three continue with an incorrect result.
Cases~11, 13, and~14 are the reverse,
where a sum should overflow.
Nimbus wraps the result, as does Prysm in Case~11,
and Lodestar computes it as a JavaScript number.
Cases~12 and~15 overflow a multiplication instead.
In Case~15, the specification multiplies the slashings total by a fixed factor
before capping it, and a large total overflows on the way.
Prysm and Nimbus multiply and wrap around,
and Lodestar divides before multiplying,
all accepting the invalid input.

\PP{State-Update Inconsistency}
In one case, a client differs from the specification
on which state field is updated, to which value, and under which condition.
In Case~16, the set \specfield{justification\_bits} bit
claims an epoch is justified
although its attestations
fall short of the two-thirds majority required for justification.
The specification decides whether to advance the justified checkpoint
from the vote count rather than from the bit,
but Prysm advances it whenever the bit is set.

\PP{Invariant Enforcement Inconsistency}
In eleven cases, clients diverge on the handling
of a broken state invariant.
Cases~17--21 give a validator an effective balance
the specification would never assign,
either above the maximum
or between the increments it rounds to.
No implementation checks the field on load,
so the state is rejected only if a later computation fails on the value.
Lodestar accepts all five.
Nimbus and Prysm reject Case~18,
where the changed balance makes them select a different proposer
than the one the block names (\specfn{compute\_proposer\_index}).
Prysm also rejects Case~19,
where adding the balance into the active-balance total overflows.
In Case~22, the value is representable, yet Lodestar still discards the
part of each effective balance between increments,
lowering its total
so the proposer receives a larger reward.

Cases~23 to~27 shorten a list that should hold one entry per validator.
In Case~23, the balance list is empty.
Lodestar reads each missing balance as zero
and produces a full-length list of zeros.
Cases~24 to~27 shorten a participation list instead.
Lighthouse processes only as many validators as the list has entries
and ignores the rest.
When the list is empty it processes none,
then divides by a total balance of zero.
Prysm and Lodestar count every missing entry
as a validator that did not participate.

\PP{Contribution of Sampling Classes.}
Each of the three value classes sampled by Algorithm~\ref{alg:sample}
reaches divergences the others do not, which is what justifies combining them.
Of the \numDivergences{} cases, 12 are reached only by boundary values
(Cases~1, 7, 9--11, 13, 17--19, 23, 24, 26),
two only by transition values (Cases~12, 20),
and six only by interior values (Cases~14, 15, 21, 22, 25, 27).
Four further cases (3--6) merge mutations drawn from more than one class.
The remaining three cases (2, 8, 16) come from type-directed fallback (Algorithm~\ref{alg:apply}),
used when no premise constrains the target field to an interval.

\PP{Practical Impact.}
Although \framework{} generates tests through proposer-side block construction,
the discovered divergences are in shared state-transition logic.
Clients run it whenever processing a block or loading a checkpoint state,
including block propagation, synchronization, and checkpoint recovery,
so these inconsistencies can propagate beyond the proposer path.

Consensus failures (Class~A) may cause clients to compute different
post-states for the same input, leading to chain splits or
stalled finality, whereas liveness failures (Class~B) remain proposer-local
but still result in invalid block production and missed block rewards.

Our generated inputs intentionally exercise adversarial boundary conditions
that are rarely encountered during normal operation.
Nevertheless, consensus clients are expected to reject invalid
inputs consistently and to compute identical post-states for the same valid
inputs. The Ethereum Foundation and client developers routinely
investigate and patch such inconsistencies, highlighting the importance
of systematically identifying them before deployment~\cite{ethspec1701}.

\subsection{RQ2: Diagnostic Power of Premise Coverage}
\label{sec:rq2}

\PP{Setup.}
We measure coverage at three levels:
premise coverage over \mech{}
(\cref{sec:coverage}),
branch coverage of \cspec,
and branch coverage of each client's state-transition code.
For \cspec and each client,
we instrument only the state-transition path,
excluding components such as BLS cryptography
and command-line modules,
using the standard coverage tool
for each implementation language.\footnote{\cspec{}: \texttt{coverage.py};
Lighthouse: \texttt{llvm-cov}; Lodestar: \texttt{c8}; Prysm: \texttt{go-bcov};
Nimbus: \texttt{gcov}; Teku: JaCoCo.
}
All measurements compare the official Capella \ctests (baseline)
against the combined suite
(baseline + \framework{}-generated tests).

\PP{Specification-Level Premise Coverage.}
After excluding \numUnfalsifiablePremises{} unfalsifiable premises
(\cref{sec:coverage}), the baseline \ctests{} falsify
\numBaselineFalsified{} of \numTotalFalsifiablePremises{} falsifiable
if-premises (\pctBaselinePremiseCoverage{}\%). None belongs to the
\numInsertedFalsifiable{} premises inserted by \mech{}. The combined
suite increases this to \numCombinedFalsified{} premises
(\pctCombinedPremiseCoverage{}\%).
This leaves \numRemaining{} premises uncovered. Of these,
\spell{\numUnattemptedPremises} are unattempted: no seed test reaches
them, leaving no seed for mutation (\cref{sec:coverage}).
The remaining \numLimitation{} fall into two categories:
\numCoordinatedLimitation{} require either more diverse seeds or
coordinated multi-field mutation, and
\spell{\numProvenanceLimitation} arise from loss of provenance information
across function calls.

\PP{Consensus-Spec Code Coverage.}
Table~\ref{tab:rq2_coverage} shows that code-level coverage changes
only marginally on \cspec{}. The combined suite adds just one line 
and two branches over the baseline.
This limited increase contrasts with the \numCtrlSpecTrumDelta{}
additional if-premises falsified by the generated tests.
Of these, \numNewFalsifiedExplicit{} correspond to conditions already present in
\cspec{}, \numNewFalsifiedInserted{} to implicit conditions added by
\mech{}, and \spell{\numNewFalsifiedSynth} to non-emptiness checks synthesized by
the elaboration pass. \cspec{} coverage captures only the first group,
yet these \numNewFalsifiedExplicit{} additional conditions account
for just one new line and two new branches.

The gap arises because \cspec{} expresses many validity
conditions implicitly. As described in \cref{sec:intro}, an invalid state
transition is signaled by raising an exception: a failed
assertion, arithmetic overflow or underflow, or an out-of-bounds
access. None introduces an explicit branch. Under
\texttt{coverage.py}'s default branch metric~\cite{consensusSpecCoverageConfig}, 
such executions contribute no additional branch coverage. A test may 
therefore exercise a new validity condition while leaving code coverage unchanged.


Even complete branch coverage cannot distinguish combinations of
independent conditions.
\cref{sec:spec} illustrates this with a simplified version of
\specfn{weigh\_justification\_and\_finalization}.
The full function additionally handles finalization, with six
independent \specplain{if}-blocks and seven feasible execution paths. The
baseline \ctests{} execute every block but reach only five paths,
whereas \framework{} reaches all seven.
%
The remaining \numNewFalsifiedInserted{} inserted premises have no counterpart in
\cspec{} and therefore cannot be measured by code-level
coverage. \mech{} makes these conditions explicit, and
premise coverage measures them directly.

\begin{table}[t]
  \centering
  \footnotesize
  \caption{Code-level coverage and case attribution
    before (Baseline) and after (Combined)
    adding \framework{}-generated tests.
    \emph{Cases} counts the RQ1 divergence cases each client is attributed,
    and \emph{Implicit} how many of those correspond to premises \mech{} inserts.
    A case attributed to several clients is counted once per client,
    so these columns exceed the \numDivergences{} cases of \cref{sec:rq1}.
}
  \label{tab:rq2_coverage}
  \setlength{\tabcolsep}{3pt}
  \renewcommand{\arraystretch}{1.15}
  \begin{tabular}{@{}l rr rr rr@{}}
    \toprule
    & \multicolumn{2}{c}{\textbf{Line / Statement}}
    & \multicolumn{2}{c}{\textbf{Branch}}
    & \multicolumn{2}{c}{\textbf{Cases}} \\
    \cmidrule(lr){2-3} \cmidrule(lr){4-5} \cmidrule(lr){6-7}
    & \multicolumn{1}{c}{Base} & \multicolumn{1}{c}{Combined}
    & \multicolumn{1}{c}{Base} & \multicolumn{1}{c}{Combined}
    & All & Implicit \\
    \midrule
    \cspec{}
      & 1,315/2,025 & 1,316/2,025
      & 152/\phantom{19,}358 & 154/\phantom{19,}358
      & \multicolumn{2}{c}{---} \\
    \midrule
    Lighthouse
      & 2,237/5,177 & 2,364/5,177
      & 177/\phantom{19,}486 & 189/\phantom{19,}486
      & 7 & 6 \\
    Lodestar
      & 3,317/4,782 & 3,431/4,782
      & 275/\phantom{19,}422 & 309/\phantom{19,}422
      & 24 & 20 \\
    Nimbus\rlap{$^a$}
      & 1,436/4,086 & 1,484/4,086
      & 804/19,270 & 907/19,270
      & 11 & 10 \\
    Prysm
      & 1,639/4,149 & 1,730/4,149
      & 188/\phantom{1}1,226 & 269/\phantom{1}1,226
      & 12 & 11 \\
    Teku
      & 2,310/5,516 & 2,342/5,516
      & 447/\phantom{1}1,258 & 480/\phantom{1}1,258
      & 0 & 0 \\
    \bottomrule
  \end{tabular}

  \smallskip
  {\scriptsize
    $^a$\,Branch denominator includes build artifacts due to limited tooling support.}
\end{table}

\PP{Client Code Coverage.}
Client code coverage exhibits the same limitation.
Across the clients, line coverage increases by only 32--127 lines and
branch coverage by 12--103 branches.
As with \cspec{}, many validity conditions do not
introduce additional branches.
For example, arithmetic overflow is handled
differently across implementations: Go, Java, and Rust (release mode)
wrap on overflow, Nim depends on compilation flags, and TypeScript has
no fixed-width integer type. None of these behaviors introduces
additional branches, even though they represent distinct validity
conditions.

%
The ``Implicit'' column of Table~\ref{tab:rq2_coverage}
quantifies this effect.
Of the \numDivergences{} divergence cases in RQ1,
\numImplicitDivergences{} cannot be found without these inserted premises.
Per client, the majority of attributed cases corresponds to inserted premises:
6 of 7 for Lighthouse, 20 of 24 for Lodestar,
10 of 11 for Nimbus, and 11 of 12 for Prysm.
Teku diverges from \cspec{} in none of the \numDivergences{} cases.
It uses the Tuweni \specplain{UInt64} library,
which provides explicit overflow checking
that \cspec{} leaves implicit.

\PP{Takeaway.}
These results establish that premise coverage
and code-level branch coverage
are complementary.
Branch coverage tracks
which control-flow paths have been exercised,
but does not capture assertion outcomes,
implicit arithmetic boundaries,
or execution paths defined by combinations of independent conditions.
This gap is not limited to Ethereum:
it arises whenever a specification expresses validity implicitly.
With a mechanized specification that makes these conditions explicit,
premise coverage can properly measure which validity conditions have been exercised.

\subsection{RQ3: Contribution of Specification Guidance}
\label{sec:rq3}

\PP{Setup.}
We compare \framework{} against two unguided generators,
\ctrlrand{} and \ctrlext{}.
Both mutate a single field per test case and draw seeds uniformly from the
\numStateTransitionTests{} state-transition pairs
that form \framework{}'s seed pool.
Neither uses premise or provenance information,
operating only on the public SSZ type structure.
\ctrlrand{} selects a field and assigns a random
type-valid value. \ctrlext{} uses the same field-selection strategy
but assigns boundary values (type extremes, neighboring values,
all-zero/all-one patterns, or boolean negation). Both retry when a
mutation leaves the input unchanged.

All three approaches run under an equal budget of
\numGeneratedTests{} generated inputs.
\framework{}'s budget includes malformed SSZ inputs,
ensuring no advantage in the number of executable test cases.

\begin{table}[t]
  \centering
  \footnotesize
  \caption{\framework{} against unguided controls under an equal budget.
    	\emph{Falsified} counts the \numCtrlTotal{} falsifiable if-premises
    	evaluated to false ($\Delta$: gain over the baseline \ctests{}).
	\emph{Cases} counts distinct divergence cases, \emph{Matched} those 
	matching \framework{}'s \numDivergences{} cases, and \emph{Class~A} 
	consensus failures under full validation.}
  \label{tab:rq3_control}
  \setlength{\tabcolsep}{4pt}
  \begin{tabular}{@{}l cc ccc@{}}
    \toprule
    & \textbf{Falsified} & \textbf{$\Delta$}
    & \textbf{Cases} & \textbf{Matched} & \textbf{Class~A} \\
    \midrule
    \ctests{} only       & \numCtrlBaseline & ---                    & ---              & ---                    & --- \\
    \quad + \ctrlrand    & \numCtrlRandom   & +\phantom{1}\numCtrlRandomDelta   & \numRandomKinds  & \phantom{1}\numRandomDivergences  & \numRandomClassA \\
    \quad + \ctrlext     & \numCtrlExtreme  & +\phantom{1}\numCtrlExtremeDelta  & \numExtremeKinds & \numExtremeDivergences & \numExtremeClassA \\
    \quad + \framework{} & \numCtrlSpecTrum & +\numCtrlSpecTrumDelta & \numDivergences  & ---                    & \numClassA \\
    \bottomrule
  \end{tabular}
\end{table}

\PP{Divergence Detection.}
Under the same testing budget, \framework{} outperforms the unguided
generators in exposing cross-client divergences. As shown in
Table~\ref{tab:rq3_control}, \ctrlrand{} exposes
\numRandomKinds{} divergence cases and \ctrlext{}
\numExtremeKinds{}, whereas \framework{} identifies \numDivergences{}.
The \emph{Matched} column counts control cases corresponding to the
same implementation fault and affected clients as a \framework{} case
(\cref{sec:rq1}). \ctrlrand{} and \ctrlext{} match
\numRandomDivergences{} and \numExtremeDivergences{} of
\framework{}'s cases, respectively. The two controls overlap only
partially, and together they match only \numControlUnion{} of
\framework{}'s \numDivergences{} cases. Under full validation,
\ctrlrand{} exposes no Class-A divergence, whereas \ctrlext{} exposes
\spell{\numExtremeClassA}, three of which match a \framework{} case.


\PP{Control-Only Divergences.}
Each control exposes \spell{\numRandomControlOnly} divergence cases not found by \framework{}.
Most arise from different seed selection: \framework{} selects \numSeeds{} seeds to maximize premise coverage,
whereas the controls sample uniformly from all \numStateTransitionTests{} state-transition seeds.
A larger seed budget would likely recover these cases (\cref{sec:discussion}).
One case in each control targets a premise already falsified by the baseline \ctests{}, which \framework{}
intentionally excludes from its target set. The remaining \ctrlext{} case mutates
\specfield{block.state\_root}, which is unchecked under
\specplain{validate\_result=False}.
Also, two inputs excluded from the counts in Table~\ref{tab:rq3_control}
expose disagreements between \cspec{} and all five clients
while the clients agree with one another. These disagreements arise from an
optimization in \cspec{} that the clients do not implement and are therefore
not considered cross-client divergences.

%

\PP{Premise Coverage.}
Premise coverage measures which conditions a test suite exercises,
rather than which faults it exposes.
\framework{} falsifies \numCtrlSpecTrumDelta{} additional premises
beyond the baseline \ctests{}, about twice as many as
\ctrlext{} (\numCtrlExtremeDelta{}) and
\ctrlrand{} (\numCtrlRandomDelta{}).
Though the metric favors \framework{} by construction,
it also shows that the implicit conditions are difficult to cover
without specification guidance.
Interestingly, \ctrlrand{} falsifies more premises than \ctrlext{},
yet exposes fewer divergences and no consensus failures.


\PP{Mutation Strategy and Field Selection.}
Both mutation strategy and field selection contribute to \framework{}'s
advantage. The controls differ only in mutation strategy, which accounts
for four additional divergence cases and all Class-A cases found by
\ctrlext{}. The remaining gap is due to field selection: without
specification guidance, a generator chooses among all 145 field paths
(65 in \specplain{BeaconState} and 80 in \specplain{BeaconBlock}),
whereas premise coverage and provenance tracking identify the
fields that affect state validity.


\subsection{RQ4: Cross-Fork Bug Reproducibility and Maintainability}
\label{sec:rq4}

Bugs in shared logic affect all fork versions
that inherit the affected functions~\cite{ethereumConsensusSpec}.
Clients support every fork version within a single codebase,
and much of the state-transition logic is shared across forks.
The specification maintainers apply fixes,
terminology changes, and refactors
across all fork versions simultaneously~\cite{ethspeccommit},
indicating that all versions remain actively maintained.
We evaluate whether the divergences identified
under the Capella configuration (RQ1)
also manifest under the Deneb configuration,
using the same specification and client versions.

Extending \mech{} from Capella to Deneb
changes +\numDenebLineChanges{} lines (about 1\%)
across \numDenebFilesChanged{} of \numDenebFilesTotal{} specification files.
The changes are local to constants, type definitions,
container structures, and fork-specific block-processing functions.
The premise count changes correspondingly,
from \numTotalIfPremises{} to \numDenebTotalIfPremises{}:
Deneb relaxes the attestation-inclusion bound,
removing five premises,
and adds one bound on blob commitments.
After adjusting the target premises
(\cref{sec:coverage}),
the remaining pipeline applies without modification,
generating \numDenebTestCases{} valid SSZ test cases
for the Deneb configuration.
Premise coverage transfers as well.
Under the same classification (\cref{sec:coverage}),
the baseline \ctests{} falsify \numDenebBaselineFalsified{} of
\numDenebTotalFalsifiablePremises{} falsifiable if-premises,
and the combined suite \numDenebCombinedFalsified{}
(\pctDenebCombinedPremiseCoverage{}\%),
a gain of \numDenebNewFalsified{} matching Capella's \numNewFalsified{}.

All \numDivergences{} divergence cases from RQ1
reproduce identically under the Deneb configuration.
Each client produces the same outcome
(acceptance, rejection, or crash)
on the corresponding Deneb-format input
as it did on the Capella-format input.
All \spell{\numDivergenceCategories} root causes
reside in functions that Deneb
inherits unchanged from Capella.

Due to the incremental structure of the specification,
if a bug exists in the shared logic, it appears in all fork versions that include it.
All \numDivergences{} cases confirm this.
Each case originated in a shared function and is reproducible in both tested fork versions.
Because the mechanization also targets shared logic,
a one-time effort on the specification core
extends to all fork versions that inherit it.
The +\numDenebLineChanges{}-line difference required to extend \mech{}
from Capella to Deneb demonstrates
that the required effort is modest
and proportional to the specification differences.

\subsection{Discussion}\label{sec:discussion}
\PP{Seeds and Mutation Breadth.}
Two factors currently limit the input space explored by
\framework{}.
The first is the seed pool.
A valid \specplain{BeaconState} must satisfy
complex cryptographic constraints,
including hash value comparisons across fields,
so \framework{} mutates \ctests{} inputs
instead of synthesizing pairs from scratch.
Then, each target premise is executed by at
most seven pre-states in our evaluation.
As shown in \cref{sec:rq3}, additional pre-states expose further divergence
cases, suggesting that a larger seed budget could improve
bug-finding ability.
The second is single-field mutation.
Premises whose falsification requires
coordinated changes to multiple fields
cannot be targeted this way,
and reaching them requires multi-field mutation.

\PP{Manual Effort.}
The mechanization is a one-time engineering effort.
\mech{} comprises \numLOC{} lines and \numTotalIfPremises{} if-premises,
written over roughly three months by a developer new to the \spectec DSL.
Once completed, all subsequent stages are automatic.
Maintenance is modest because Ethereum specifications evolve
incrementally across forks.
For example, extending \mech{} from Capella to Deneb required only \numDenebLineChanges{}
changed lines (\cref{sec:rq4}). 
Classifying the \numTotalIfPremises{} if-premises into \numTautologyPremises{}
tautologies, \numClosingBranchPremises{} closing branches,
and \numTotalFalsifiablePremises{} falsifiable premises (\cref{sec:coverage}) took
roughly two days. The process is largely mechanical, and could be partly automated with
structural checks and SMT-based tautology detection for a larger specification.
An error in this classification is also bounded in consequence,
because \mech{} guides generation rather than deciding correctness (\cref{sec:diff_testing}).

\PP{Non-Target Premise Classification.}
Excluding non-targets is conservative.
Misclassifying a target premise as a non-target
only makes the generator skip a valid target,
which may reduce coverage but does not produce false positives.

\PP{Generality.}
Once the mechanization is in place,
the rest of the pipeline is domain-agnostic.
It applies to any domain with several independent implementations of a
specification whose validity is left implicit.

\subsection{Threats to Validity}\label{sec:validity}

\PP{Internal Validity.}
Our evaluation relies on two reference artifacts:
\cspec{} for fault attribution and \mech{}
for test generation. If \cspec{} is incorrect, fault
attribution may be affected. Such errors would appear as
inputs on which \cspec{} alone disagrees with all
clients. Our controls exposed two such cases (\cref{sec:rq3}), both
caused by an optimization in \cspec{} rather than
client faults. Faults shared by \cspec{} and a subset
of clients would not be detected in this way. For
\mech{}, incorrect premises may direct test
generation toward unintended boundaries. We mitigate this
threat by deriving premises directly from \cspec{}
and validating the generated interpreter against all 910 official
reference tests.

\PP{Construct Validity.}
Branch coverage tools vary in capability
across the five implementation languages,
which may affect absolute coverage values.
The \code{go-bcov} tool we use for Prysm may miss short-circuit evaluation,
and Nimbus's branch denominator includes build artifacts
due to limited \code{gcov} support for Nim.
These differences affect cross-client comparisons,
but for each client, the baseline and combined suites use identical instrumentation.

\PP{External Validity.}
We evaluate \framework{} on one protocol version (Capella)
with cross-fork reproduction on Deneb.
The specification defines seven fork versions.
While RQ4 shows that bugs in the shared logic
reproduce across forks,
\framework{} targets the state-transition function
and does not cover fork-choice rules,
light-client protocols, or the execution layer.
We test the five major actively maintained clients,
excluding minor and experimental ones.

\section{Related Work}
\PP{Specification-Guided Testing.}
JEST~\cite{jest} mechanizes the ECMAScript specification
and generates JavaScript programs to differentially test browser engines.
WEST~\cite{west} generates Wasm programs from a mechanized Wasm specification
using the \spectec~\cite{spectec} toolchain.
Closest to our work,
Lee et al.~\cite{p4ntt} mechanize the P4 type system in P4-SpecTec,
define \textit{dangling coverage} over the premises
whose violation makes a program ill-typed,
and mutate well-typed seed programs to cover them.
RFCcert~\cite{rfccert} extracts validation rules from RFC prose
and assembles certificates by dynamic symbolic execution.
All four operate on specifications that state which inputs are valid,
in the form of ECMAScript early errors, Wasm validation rules,
P4 typing rules, and X.509 certificate requirements.
Since \cspec{} does not specify validation conditions and
these conditions exist solely as Python control flow,
\framework{} recovers them as if-premises,
enabling the definition of premise coverage.

\PP{Differential Fuzzing for Blockchain Consensus.}
\textit{beacon-fuzz}~\cite{beaconfuzz} is a structure-aware differential fuzzer
that generates \specplain{BeaconState} and \specplain{BeaconBlock} pairs
and compares the outcomes of the Ethereum consensus clients.
Fluffy~\cite{yang2021} targets the execution layer,
mutating transaction order and semantics
to expose divergences across execution clients.
Forky~\cite{kim2025fork} constructs valid fork scenarios
to stress the fork-resolution logic.
Loki~\cite{ma2023loki} captures live node states
and mutates them via type-specific rules
to explore high-dimensional input spaces.
All four treat clients as black boxes
and select mutation targets without referring
to the specification's validity conditions,
so they cannot identify
which semantic boundaries remain untested.

\PP{Differential Testing beyond Blockchain.}
Differential testing of multiple implementations
conforming to a shared specification
has been applied to JVMs~\cite{classfuzz,classming},
SSL/TLS libraries~\cite{mucert},
EVM bytecode interpreters~\cite{fu2024evmfuzz},
and Wasm runtimes~\cite{wadiff}.
These tools generate inputs via syntactic mutation
or code-coverage guidance.
Our mechanized specification instead exposes validity conditions
that the client's branching structure does not express,
so code-level coverage cannot detect them (\cref{sec:rq2}).

\PP{Formal Verification of Consensus Protocols.}
Gasper~\cite{buterin2020combining} has been verified for finality
using Coq~\cite{verifying-Gasper},
and model-checking approaches have targeted
individual Beacon Chain components
such as validator exits~\cite{rashid2023voluntary}
and reward mechanisms~\cite{rashid2025formal}.
These efforts prove properties over abstract models
but do not test production code.
Our framework is complementary,
targeting implementation-level conformance across five clients.

\section{Conclusion}\label{sec:conclusion}
We present \framework{},
a specification-guided differential testing framework
for Ethereum consensus clients.
By mechanizing the consensus specification into explicit if-premises,
defining premise coverage over them,
and generating tests that target under-exercised conditions,
\framework{} improves the premise coverage of falsifiable premises
from \pctBaselinePremiseCoverage{}\% of the official test suite to \pctCombinedPremiseCoverage{}\%.
Applied to five major clients,
it exposes \numDivergences{} cross-client divergence cases,
including consensus failures caused by lossy \specplain{uint64} decoding
and client crashes on empty inputs.
All \numDivergences{} cases reproduce across fork versions,
and extending the mechanized specification to a new fork
requires effort proportional to the specification difference.
These results show that cross-client divergences are most likely
to arise at validity boundaries that \cspec{} leaves implicit
and the \ctests{} fail to systematically exercise.

\begin{acks}
This work was supported by the National Research Foundation of Korea (NRF)
(2021R1A5A1021944, 2022R1A2C2003660, and 2026-25549584)
and the Institute of Information \& Communications Technology Planning \& Evaluation (IITP)
grant funded by the Korea government (MSIT) (2022-0-00688 and 2024-00337703).
\end{acks}
\section*{Data Availability}
\framework{} is available at~\cite{spectrum-artifact-2026}. It contains \mech{},
the test generator, the differential testing harness, and the
scripts for reproducing the experiments.

\balance
\bibliographystyle{ACM-Reference-Format}
\bibliography{ref}

\end{document}